\documentclass{CSML}

\def\dOi{12(1:2)2016}
\lmcsheading%
{\dOi}
{1--34}
{}
{}
{Feb.~\phantom02, 2015}
{Mar.~\phantom02, 2016}
{}

\ACMCCS{[{\bf Theory of computation}]: Logic---Proof theory\,/\,Linear
  logic; Computational complexity and cryptography---Proof complexity}
\usepackage{amssymb,amsthm,amsmath,hyperref}
\usepackage{microtype}
\usepackage[T1]{fontenc}
\usepackage{textcomp}
\usepackage{mathtools}

\usepackage{LMCS_macros,willemtools}

\longJournalNames
\let\capsabbrev=\uppercase

\title{Proof equivalence in \capsabbrev{mll} is \capsabbrev{pspace}-complete}

\author[W.~Heijltjes]{Willem Heijltjes\rsuper a}
\address{{\lsuper a}University of Bath}
\email{w.b.heijltjes@bath.ac.uk}

\author[R.~Houston]{Robin Houston\rsuper b}
\address{{\lsuper b}Kiln Enterprises}
\email{robin@kiln.it}

\begin{document}






\begin{abstract}
\capsabbrev{mll} proof equivalence is the problem of deciding whether two proofs in multiplicative linear logic are related by a series of inference permutations.
It is also known as the word problem for $*$-autonomous categories.
Previous work has shown the problem to be equivalent to a rewiring problem on proof nets, which are not canonical for full \capsabbrev{mll} due to the presence of the two units.
Drawing from recent work on reconfiguration problems, in this paper it is shown that \capsabbrev{mll} proof equivalence is \capsabbrev{pspace}-complete, using a reduction from Nondeterministic Constraint Logic.
An important consequence of the result is that the existence of a satisfactory notion of proof nets for \capsabbrev{mll} with units is ruled out (under current complexity assumptions).
The \capsabbrev{pspace}-hardness result extends to equivalence of normal forms in \capsabbrev{mell} without units, where the weakening rule for the exponentials induces a similar rewiring problem.
\end{abstract}

\maketitle


\section{Introduction}
\noindent
Sequent calculus was originally introduced by Gentzen as a way to better understand
the properties of natural deduction -- in particular the eliminability of cut. In this
view, a sequent calculus proof gives a recipe for the construction of a natural deduction
proof.
Linear logic was formulated as a sequent calculus system, and doesn't have a corresponding
natural deduction in the 
traditional sense. The role of natural deduction is taken instead by
proof nets. Each sequent calculus proof gives rise to a proof net. Furthermore, at least in
the case of multiplicative linear logic without units, proof nets are canonical in the sense
that two sequent calculus proofs give rise to the same proof net if and only if they are
equivalent.

Proof nets do not work so well when the logical units 1 and $\bot$ are included. Although
proof nets can still be defined in this case, they are no longer canonical. Firstly the
translation from proofs to proof nets is not canonical, so a proof may be translated
to several possible proof nets with no privileged choice. Also, equivalent proofs are not
in general translated to the same possible proof nets. So proof nets lose their principal
advantage, compared to proofs, of being canonical, and are subject to a non-trivial
equivalence relation.

Equivalence of proofs (in the sense we are interested in here) was first considered
by Lambek \cite{Lambek-1968}, who introduced the idea of category-theoretic logic and used
it to define a natural notion of equivalence of proofs.

Proof nets were introduced at the birth of linear logic by Girard \cite{Girard-1987}, and
the correctness criterion was simplified by Danos and Regnier \cite{Danos-Regnier-1989}.
These proof nets did not include the logical units, and are canonical for multiplicative linear logic.
Canonical proof nets also exist for larger fragments of linear logic, such as the combined multiplicative-additive fragment without units \cite{Hughes-vanGlabbeek-2005}, and the additive fragment, including the additive units \cite{Heijltjes-2011}.

Proof nets incorporating units were defined by Blute, Cockett, Seely and Trimble \cite{Trimble-1994,Blute-Cockett-Seely-Trimble-1996}, though these proof nets cease to be canonical. The `joker games' of Koh and Ong \cite{Koh-Ong-1999}, for \capsabbrev{mll} and its intuitionistic variant \capsabbrev{imll}, have a very similar structure to proof nets, and include units. Another explicit treatment of proof nets for \capsabbrev{mll} with units is given by Lamarche and Stra\ss burger \cite{Lamarche-Strassburger-2006}, and later a particularly simple and elegant version by Hughes \cite{Hughes-2012-categories,Hughes-2012-nets}. However, none of these notions gives proof nets that are canonical.


The difficulty with the multiplicative units is that each instance of $\bot$, the unit to the par, must be attached to the main proof structure, while there isn't always a canonical point where to attach it. In the sequent calculus, $\bot$-instances are introduced by a weakening rule, which permutes with many (but not all) other rules. In proof nets, $\bot$-instances are commonly attached by edges called jumps, which may be \emph{rewired} via local graph-rewrite rules \cite{Trimble-1994,Blute-Cockett-Seely-Trimble-1996}. For intuitionistic \capsabbrev{mll} rewiring may be directed towards the unique conclusion, to yield canonical forms.\footnote{The canonical nets for \capsabbrev{imll} described here appear to be folklore; established proof nets for intuitionistic linear logic, called essential nets \cite{Lamarche-2008,Murawski-Ong-2003}, (non-canonically) attach jumps to the leaves of the formula tree, for easier composition.}
For classical MLL with units, however, there is not an obvious preferred point to attach jumps. And this problem extends to multiplicative--exponential linear logic (\capsabbrev{mell}) without units, where a formula $?A$ may be introduced via a weakening rule with similar permutability to the $\bot$-introduction rule.


The idea of the present work is to look at proof equivalence through the lens of computational
complexity, and use this to settle the question of whether there can be canonical proof nets
for \capsabbrev{mll} with units. The canonical proof nets for \capsabbrev{mll} without units
give an efficient decision procedure for proof equivalence: to decide whether two such proofs
are equivalent, it is sufficient to translate both to proof nets and compare the proof nets
for equality.
We show that the corresponding decision procedure for \capsabbrev{mll} \emph{with} units is
\capsabbrev{pspace}-complete, which is generally supposed to preclude the existence of a
polynomial-time algorithm for this problem. So there can be no canonical proof nets in the
usual sense: if the proof nets are syntactically equal just when their corresponding proofs
are equivalent, then the translation from proofs to proof nets must be intractable.


In \capsabbrev{mell}, even without units, the unit $\bot$ may be emulated via carefully chosen formulae $?A$. Consequently, our result means that equivalence of cut-free MELL proofs is \capsabbrev{pspace}-hard. This is in sharp contrast with many intuitionistic calculi such as the simply typed lambda-calculus, where normal forms are unique. However, it does not impact the complexity of proof equivalence for \capsabbrev{mell} in general, which is dominated by cut-elimination (\capsabbrev{mell} encodes the simply-typed lambda-calculus, which is not elementary recursive \cite{Statman-1977}).



Given our results for classical linear logic, which effectively rule out canonical forms for \capsabbrev{mll} and \capsabbrev{mell}, it becomes tempting to consider possible alternatives. One is intuitionistic linear logic, mentioned above. Another is Melli\`es's game-semantics inspired \emph{tensorial logic} \cite{Mellies-2012}, with a semantics in \emph{dialogue categories}. Tensorial logic rejects double-negation elimination, and embeds linear logic via a double-negation translation; the presence of explicit negations in formulae renders jumps immobile, so that the re-wiring problem for the unit does not occur. A further alternative is the \emph{polarised} fragment of linear logic, which has canonical proof nets \cite{Laurent-1999}; here, the rewiring of jumps is blocked by the presence of exponentials rather than negations. Finally, in a direction aimed more at proof search, \emph{focusing} retains classical linear logic but weakens the notion of proof equivalence. This gives canonical representations in \emph{focused proof nets} \cite{Andreoli-Maieli-1999} and \emph{multi-focused proofs} \cite{Chaudhuri-Miller-Saurin-2008}.


Our proof of \capsabbrev{pspace}-completeness uses the \emph{nondeterministic constraint logic} of \cite{Hearn-Demaine-2005,Demaine-Hearn-2008,GamesPuzzlesAndComputation}. This is a graph rewriting
system introduced as a tool for use in complexity proofs, originally for games and puzzles.

It is not uncommon for an \capsabbrev{np}-complete problem to have an associated reconfiguration problem that is \capsabbrev{pspace}-complete \cite{ReconfigurationProblems}; an example of this is \capsabbrev{sat}-reconfiguration. \capsabbrev{mll} proof equivalence may be regarded as the reconfiguration problem associated with \capsabbrev{mll} proof search, which is \capsabbrev{np}-complete \cite{Kanovich-1992,Lincoln-Winkler-1994}.


This paper is an extended version of \cite{Heijltjes-Houston-2014}.



\section{\texorpdfstring{\protect\capsabbrev{mll}}{MLL}}

In this section, we will give a brief introduction to multiplicative linear logic (\capsabbrev{mll}). For simplicity of exposition, we will work in the unit-only fragment of \capsabbrev{mll}. The formulae of this fragment are given by the following grammar.
\setMidspace{8pt}
\[
	A,B,C \Coloneq \bot \Mid 1 \Mid A\pr B \Mid A\tn B
\]
The connectives tensor ($\tn$) and par ($\pr$) will be considered up to associativity, and \emph{duality}, indicated $\dual A$, is via DeMorgan.
A \emph{sequent} $\Gamma, \Delta,\Lambda$ will be a multiset of formulae.
To distinguish different occurrences of a formula within a sequent, the connectives and units in a sequent will be \emph{named} with distinct elements from an arbitrary set of names, e.g.:\
\[
	\named a1\named b\pr\named c1~,~\named d\bot\named e\tn\named f\bot~.
\]
This simple technical device is introduced for a dual purpose. Firstly, it distinguishes proofs that are equal up to a symmetric exchange of formulae, such as the \emph{identity} and the \emph{twist} proof of the above sequent, while the sequent rules can be given using standard multiset sequents, i.e.\ without the need for explicit \emph{exchange} rules. Secondly, it creates an easy way to extract proof nets, as graphs that use the names of connectives and units as vertices.

Within a proof, the names of units and connectives are preserved through inferences. Where convenient, we may leave names implicit.

\begin{figure}
\[
		\MLLrule b
\qquad	\MLLrule 1
\qquad	\MLLrule p
\qquad	\MLLrule t
\qquad \vc{\rule{.5pt}{30pt}}
\qquad	\MLLrule a
\qquad	\MLLrule c
\]
\caption{Inference rules for unit-only \capsabbrev{mll}}
\label{fig:MLL}
\end{figure}

\begin{figure}
\[
\begin{array}{c}
\\
	\MLLelim{a0}\quad\rightsquigarrow\quad\MLLelim{a1}
\qquad\qquad\qquad
	\MLLelim{a2}\quad\rightsquigarrow\quad\MLLelim{a3}
\\[5pt]
\end{array}
\]
\caption{Identity-elimination transformations}
\label{fig:id-elim}
\bigskip
\[
\begin{array}{rcl}
\\[5pt]
	\MLLelim{c0} & \rightsquigarrow & \Gamma \\ \\[10pt]
	\MLLelim{c1} & \rightsquigarrow & \MLLelim{c2}
\\[5pt]
\end{array}
\]
\caption{Cut-elimination transformations}
\label{fig:cut-elim}
\bigskip
\[
\begin{array}{c}
\\[5pt]
	\MLLperm{bb1}\quad\sim\quad\MLLperm{bb2}
	\qquad\qquad
	\MLLperm{bp1}\quad\sim\quad\MLLperm{bp2}
\\ \\[10pt]
	\MLLperm{bt2}\quad\sim\quad\MLLperm{bt1}\quad\sim\quad\MLLperm{bt3}
\\ \\[10pt]
  \begin{array}{rcl}
	\MLLperm{pp1} & \sim & \MLLperm{pp2}
  \\ \\[10pt]
	\MLLperm{pt1} & \sim & \MLLperm{pt2}
  \\ \\[10pt]
	\MLLperm{tt1} & \sim & \MLLperm{tt2}
  \end{array}
\\[5pt]
\end{array}
\]
\caption{Permutations}
\label{fig:permutations}
\end{figure}

A one-sided sequent calculus for unit-only \capsabbrev{mll} is given in Figure~\ref{fig:MLL}. The rules for identity ($\mathsf{ax}$) and cut ($\mathsf{cut}$) are admissible, and are eliminated via the identity-elimination and cut-elimination transformations displayed in Figure~\ref{fig:id-elim} and Figure~\ref{fig:cut-elim} respectively. The cut-elimination process in addition makes essential use of \emph{permutations} of inference rules, displayed in Figure~\ref{fig:permutations}, to match up a cut-rule with the two inferences that introduce both cut-formulae. Note that the symmetric variants of various transformations have been omitted from the figures, as well as permutations involving the cut-rule, which are similar to those of the tensor rule.

The proof system we shall be working with is that consisting of just the introduction rules for the two units and the two connectives, the four rules $(\bot)$, $(\1)$, $(\pr)$, and $(\tn)$ of Figure~\ref{fig:MLL}. The problem we consider is the following.

\begin{defi}[\capsabbrev{mll} proof equivalence]
\label{def:equivalence}
\emph{Equivalence} $(\perm)$ of proofs in (unit-only, cut-free, identity-free) multiplicative linear logic is the congruence generated by the permutations given in Figure~\ref{fig:permutations}. \emph{\capsabbrev{mll} proof equivalence} is the problem of deciding whether two given proofs are equivalent.
\end{defi}


\subsection{Star-autonomous categories}

The permutations of sequent proofs are exactly the identifications imposed by the categorical semantics of \capsabbrev{mll}, star-autonomous categories \cite{Barr-1991} (and semi-star-autonomous categories \cite{Houston-2008,Heijltjes-Strassburger} for \capsabbrev{mll} without units). Proof equivalence for \capsabbrev{mll} is equivalent to the \emph{word problem} for star-autonomous categories: the problem of deciding whether two term representations denote the same morphism in any star-autonomous category.


\section{Proof nets}
\enlargethispage{\baselineskip}

For \capsabbrev{mll} without units, Girard's original proof nets \cite{Girard-1987} are canonical: two proofs translate into the same proof net if and only if they are equivalent. Since the translation from proofs to proof nets and the syntactic comparison of proof nets are both effectively computable (linear-time), proof nets provide an effective solution to the proof equivalence problem for \capsabbrev{mll} without units.

There have been several proposals for (non-canonical) multiplicative proof nets with units \cite{Blute-Cockett-Seely-Trimble-1996, Koh-Ong-1999, Lamarche-Strassburger-2006, Hughes-2012-nets}, each providing a different take on the proof equivalence problem for \capsabbrev{mll}.
We will use essentially the formulation by Hughes \cite{Hughes-2012-nets}. A proof net is a sequent, seen as a forest of formula trees, for which a set of \emph{links} or \emph{jumps} attaches the instances of $\bot$ to other vertices (occurrences of connectives and units) in the forest.
Rewiring is the re-attachment of one jump, either as a small-step relation that moves jumps only to neighbouring vertices, or as a big-step relation that moves a jump anywhere in the forest. Hughes relates both flavours of rewiring, which allows us to switch between the two at will. In addition, we may at any point impose the restriction that jumps connect only to occurrences of the unit $1$.

In this section we will discuss how this notion of proof net arises from the sequent calculus, and introduce a compact notation for it.

\begin{defi}[Proof nets]
\label{def:proof nets}
For a sequent $\Gamma$,
\begin{itemize}

	\item
	a \emph{linking} $\links$ is a function from the names of $\bot$-subformulae in $\Gamma$ to the names of $\Gamma$;

	\item
	a \emph{switching graph} for $\Gamma$ and $\links$ is an undirected graph over the names of $\Gamma$ with:
	\begin{itemize}
		\item for every subformula $\named aA\named c\tn\named bB$ the edges $\edge ca$ and $\edge cb$,
		\item for every subformula $\named aA\named c\pr\named bB$ either the edge $\edge ca$ or the edge $\edge cb$, and
		\item for every subformula $\named a\bot$ the edge $\edge a{\links(a)}$;
	\end{itemize}
	
	\item
	a linking $\links$ is \emph{correct} if every switching graph is acyclic and connected;

 	\item
	a \emph{proof net} $(\Gamma,\links)$ consists of the sequent $\Gamma$ and a correct linking $\links$ for $\Gamma$.

\end{itemize}
\end{defi}

An example proof net is given in Figure~\ref{fig:example net}. The edges from a $\pr$-node, which are subject to being switched, are drawn as dashed lines, while edges from a $\tn$-node are solid lines. An edge $\edge a{\links(a)}$ in a proof net or switching graph is called a \emph{link} or a \emph{jump}, and may also be indicated $\jump a{\links(a)}$, to match the illustrations. Note that unlike the axiom links of proof nets without units, a jump may connect to any connective or unit, even another $\bot$-instance, and multiple jumps may point to the same position. A path between vertices $a$ and $b$ in a switching graph is indicated as $\path ab$.

\begin{figure}
\[
	\examplenet
\]
\caption{An example proof net}
\label{fig:example net}
\end{figure}

The present notion of proof net can be seen as a direct interpretation of the sequent calculus, in the following way. As there are inequivalent proofs of the sequent $\bot\tn\bot,1\pr1$, some way of attaching the $\bot$-formulae to the remainder of the proof net appears necessary. The introduction rule for $\bot$, in Figure~\ref{fig:MLL}, attaches a $\bot$-formula to a sequent---but sequents are not explicit in proof nets. The notion of a jump, which attaches the $\bot$ to an arbitrary subformula in the sequent $\Gamma$, is then a natural generalisation of the axiom links of unit-free \capsabbrev{mll} proof nets. This gives the following translation from proofs to proof nets.

\begin{defi}
\label{def:proofs to nets}
The relation $(\toNet)$ interprets a proof $\Pi$ by a linking $\links$ as follows: $\Pi\toNet\links$ if for each $\named a\bot$ in $\Pi$, $\links(a)$ is a name in the context $\Gamma$ of the inference introducing $\named a\bot$:
\[
	\infer[\MLLlabel b]{\Gamma,\named a\bot}{\Gamma}~.
\]
\end{defi}

An example is given in Figure~\ref{fig:example trans}, where a proof is interpreted as the proof net of Figure~\ref{fig:example net}, by indicating the chosen jumps in the conclusion of each $\bot$-inference. (In the example, to trace formulae through inferences, the five $\bot$-formulae are kept in order from left to right, while the three $1$-formulae have explicit names.)

\begin{figure}
\[
	\exampletrans
\]
\caption{Interpreting a proof as the proof net of Figure~\ref{fig:example net}}
\label{fig:example trans}
\end{figure}

\begin{prop}[\cite{Danos-Regnier-1989,Lamarche-Strassburger-2006}]
\label{prop:correctness and sequentialisation}
If $\Pi\toNet\links$ and $\Pi$ has conclusion $\Gamma$, then $(\Gamma,\links)$ is a proof net. For any proof net $(\Gamma,\links)$ there is a proof $\Pi$ of $\Gamma$ such that $\Pi\toNet\links$ (sequentialisation).
\end{prop}

\begin{rem}
The proof nets of \cite{Blute-Cockett-Seely-Trimble-1996} may appear to be formulated in quite a distinct way from our presentation, in particular in that their jumps from $\bot$-vertices connect to edges, rather than vertices. The difference is only superficial: in their proof nets, vertices represent inferences, and edges represent formulae; in both presentations, therefore, jumps may be seen to attach to formulae. This way of attaching jumps (to formulae, rather than inferences) is the natural choice also in their setting, which is motivated categorically, where $\bot$-introduction corresponds to the isomorphism $A\to A\pr\bot$.
\end{rem}

\subsection{Rewiring}

The use of proof nets factors out the bottom three permutations in Figure~\ref{fig:permutations}, $(\pr-\pr$), $(\tn-\pr)$, and $(\tn-\tn)$. The remaining permutations, involving the $\bot$-introduction rule, impose an equivalence on proof nets, consisting of the \emph{rewiring} of jumps from one target to another. This is defined as follows.

\begin{defi}
\label{def:proof net equivalence}
A \emph{rewiring} $(\perm*)$ between proof nets changes the target of exactly one jump.
\emph{Equivalence} $(\perm)$ of proof nets over a sequent $\Gamma$ is the equivalence generated by rewiring.
\end{defi}

We will write $\links[\jumpsto ab]$ for the linking where $\jump ab$, and $\jump c{\links(c)}$ for any $\named c\bot$ other than $a$. Then in a rewiring $\links\perm*\links[\jumpsto ab]$, the jump being rewired is that from $a$.

The above definition, which is due to Hughes~\cite{Hughes-2012-nets}, gives the ``big-step'' rewrite relation for equivalence. In turn, this rewiring relation is generated by smaller local steps, which correspond more closely to the permutations in the sequent calculus. These smaller steps are illustrated in Figure~\ref{fig:rewiring}. In the top row, a jump can be rewired across another jump and across a tensor without breaking correctness; these rewiring steps can be seen to correspond to $(\bot-\bot)$ and $(\tn-\bot)$ permutations. In the bottom row, rewiring a jump into the par, from $a$ or $b$, only preserves correctness if the jump can be rewired between $a$ and $b$ independently of the par. Intuitively, this corresponds to the $(\pr-\bot)$ permutation in the following way: the positions that a jump may rewire to correspond to the context of the rule introducing the $\bot$; then to have the premise of the $\pr$-introduction as its context, the jump should be able to rewire to both $a$ and $b$.

That the rewiring relation is generated by the small-step relation in Figure~\ref{fig:rewiring} is a main theorem in \cite{Hughes-2012-nets}.

\begin{prop}[\cite{Hughes-2012-nets}]
Rewiring $(\perm*)$ is generated by the local steps in Figure~\ref{fig:rewiring}.
\end{prop}

\begin{figure}
\[
\begin{array}{cccrccc}
	\PN b1 & \sim & \PN b2 &                          \PN t1 \quad \sim  & \PN t2 & \sim & \PN t3
\\ \\ \\
	\PN p1 & \sim & \PN p2 & \sim \quad \PN p3 \qquad\text{only if}\quad & \PN q1 & \sim & \PN q2
\end{array}
\]
\caption{Rewiring steps}
\label{fig:rewiring}
\end{figure}

A further refinement in the proof nets introduced by Hughes \cite{Hughes-2012-nets}, is that jumps may be restricted to target only $1$-formulae. This will be convenient for the compact diagrammatic notation introduced below, and unless otherwise indicated, we will assume that proof nets are of this form.
 

\begin{prop}[\cite{Hughes-2012-nets}]
Any proof net is equivalent to one where the codomain of $\links$ is restricted to occurrences of the formula $1$.
\end{prop}

A main technical advantage of this restriction is that cut-elimination can be performed in a one-shot operation, by path composition over jumps. Then, without the need to define cut-elimination by local transformations, proof nets may omit an explicit representation of cuts, as the present definition does.

Finally, the following proposition states that equivalence is preserved when translating between proofs and proof nets, which means that \capsabbrev{mll} proof equivalence is equivalent to the problem of deciding equivalence of proof nets.

\begin{prop}[\cite{Blute-Cockett-Seely-Trimble-1996,Hughes-2012-nets}]
\label{prop:proof nets work}
If $\Pi\toNet\links$ and $\Pi'\toNet\links'$, then $\Pi\perm\Pi'$ iff $\links\perm\links'$.
\end{prop}


\subsection{Diagram notation}

To concisely represent larger sequents and proof nets, we will introduce a compact diagrammatic notation.
Briefly, the units $1$ and $\bot$ are represented by a circle~($\circ$) and a disc~($\bullet$) respectively, a tensor $(\tn)$ will be indicated by an edge, and a par $(\pr)$ will be indicated by a box.
For example, the diagram below left represents the sequent below right.
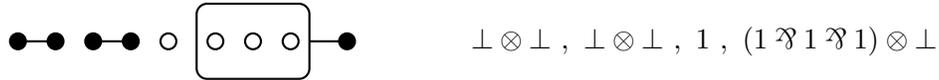
\[
\begin{array}{c}
\vc{\begin{tikzpicture}[x=5mm,y=-5mm,octo]
	\draw (0,0) node[bullet] (a) {} -- (1,0) node[bullet] (b) {} (2,0) node[bullet] (c) {} -- (3,0) node[bullet] (d) {};
	\node[circ] (1) at (4,0) {};
	\node[circ] (2) at (5.25,0) {}; \node[circ] (3) at (6.25,0) {}; \node[circ] (4) at (7.25,0) {};
	\draw[rounded corners] (4.75,-1) rectangle (7.75,1);
	\draw (7.75,0) -- (8.75,0) node[bullet] (e) {};
\end{tikzpicture}}
\qquad\qquad
	\bot\tn\bot~,~\bot\tn\bot~,~\1~,~(\1\pr\1\pr\1)\tn\bot
\end{array}
\]
The salient features of a diagram are the connectedness by $\tn$-edges and containment within a $\pr$-box. Respecting these, we will lay out a diagram in the plane in any way that is convenient, while ensuring (of course) that the layout is kept consistent when comparing different proof nets over the same sequent. More formally, then, sequents and formulae are interpreted as diagrams as follows:
\[
  1\mathrel\Rightarrow\circ, \quad  \bot\mathrel\Rightarrow\bullet, \quad \text{and if}\quad
	A_i~\Rightarrow~
	  \vc{\begin{tikzpicture}[octo]
	    \node[draw,rounded corners] {$A_i$};
	  \end{tikzpicture}}
  \quad\text{then}
\]
\[
\begin{array}{@{}rcll@{}}
	A_1\tn A_2\tn\dotso\tn A_n 
	& \Rightarrow &
	  \vc{\begin{tikzpicture}[octo]
	    \node[draw,rounded corners] (1) at (0,0) {$A_1$};
	    \node[draw,rounded corners] (2) at (1,0) {$A_2$};
	    \node[draw,rounded corners] (n) at (3,0) {$A_n$};
	    \node at (2,0) {$\dots$};
	    \draw (1)--(2)--(1.6,0) (2.4,0)--(n);
	    \path[use as bounding box] (-.5,-.5) rectangle (3.5,.5);
	  \end{tikzpicture}}
	& \text{where no $A_i$ is a $\tn$-formula}
\\ \\ 
	A_1\pr A_2\pr\dotso\pr A_n
	& \Rightarrow &
	  \vc{\begin{tikzpicture}[octo]
	    \node[draw,rounded corners] at (0,0) {$A_1$};
	    \node[draw,rounded corners] at (1,0) {$A_2$};
	    \node[draw,rounded corners] at (3,0) {$A_n$};
	    \node at (2,0) {$\dots$};
	    \draw[rounded corners] (-.5,-.5) rectangle (3.5,.5);
	  \end{tikzpicture}}
	& \text{where no $A_i$ is a $\pr$-formula}
\\ \\
	A_1, A_2,\dotsc, A_n
	& \Rightarrow &
	  \vc{\begin{tikzpicture}[octo]
	    \node[draw,rounded corners] at (0,0) {$A_1$};
	    \node[draw,rounded corners] at (1,0) {$A_2$};
	    \node[draw,rounded corners] at (3,0) {$A_n$};
	    \node at (2,0) {$\dots$};
	    \path[use as bounding box] (-.5,-.5) rectangle (3.5,.5);
	  \end{tikzpicture}}
	& 
\end{array}
\]
In the diagrammatic notation, as in the regular notation before, the jumps of a proof net are added to the sequent as coloured arrows. Per the restriction mentioned previously, jumps always connect a $\bot$-formula to a $1$-formula. The example below gives a proof net for the above sequent, presented in both notational styles.
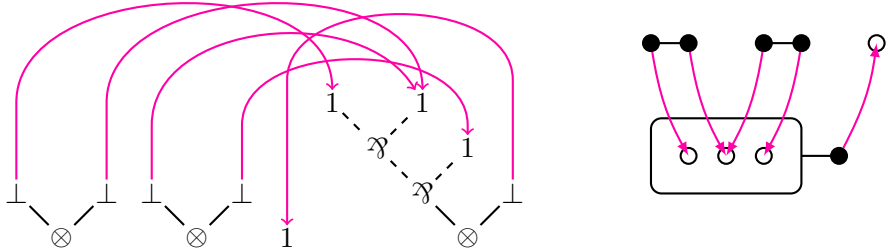
\[
\vc{\begin{tikzpicture}[x=6mm,y=6mm,inner sep=1pt]
	\node (a) at (0,1) {$\bot$};
	\node (b) at (1,0) {$\tn$};
	\node (c) at (2,1) {$\bot$};
	\node (d) at (3,1) {$\bot$};
	\node (e) at (4,0) {$\tn$};
	\node (f) at (5,1) {$\bot$};
	\node (g) at (6,0) {$1$};
	\node (h) at (7,3) {$1$};
	\node (i) at (8,2) {$\pr$};
	\node (j) at (9,1) {$\pr$};
	\node (k) at (9,3) {$1$};
	\node (l) at (10,0) {$\tn$};
	\node (m) at (10,2) {$1$};
	\node (n) at (11,1) {$\bot$};
	\node at (0,-.5) {};
	\draw[thick] (a)--(b)--(c) (d)--(e)--(f) (j)--(l)--(n);
	\draw[thick,dashed] (h)--(i)--(k) (i)--(j)--(m);
	\draw[jump] (a) -- +(0,2) to[out=90,in=90] (h);
	\draw[jump] (c) -- +(0,2) to[out=90,in=90] (k);
	\draw[jump] (d) -- +(0,1.5) to[out=90,in=120] (k);
	\draw[jump] (f) -- +(0,1.5) to[out=90,in=90] ($(m)+(0,0.5)$) -- (m);
	\draw[jump] (n) -- +(0,2.5) to[out=90,in=90] ($(g)+(0,3.5)$) -- (g);
\end{tikzpicture}}
\qquad\qquad
\vc{\begin{tikzpicture}[x=5mm,y=-5mm,octo]
	\node[circ] (1) at (6,0) {};
	\draw (0,0) node[bullet] (a) {} -- (1,0) node[bullet] (b) {} (3,0) node[bullet] (c) {} -- (4,0) node[bullet] (d) {};
	\node[circ] (2) at (1,3) {}; \node[circ] (3) at (2,3) {}; \node[circ] (4) at (3,3) {};
	\draw[rounded corners] (0,2) rectangle (4,4);
	\draw (4,3) -- (5,3) node[bullet] (e) {};
	\begin{scope}[thepink,->]
			\draw[bend right=10] (a) to (2.center);
			\draw[bend right=10] (b) to (3.center);
			\draw[bend left=10]  (c) to (3.center);
			\draw[bend left=10]  (d) to (4.center);
			\draw[bend right=10] (e) to (1.center);
	\end{scope}
\end{tikzpicture}}
\]
The above examples illustrate how the diagrammatic notation replaces the connectives by geometric features to allow for more convenient placement in the plane. The diagrams preserve the connectedness of non-switched edges, while switched components are placed within a box, so that a switching graph is obtained by replacing each box by an edge to exactly one of its components.

To lay out a sequent as a diagram in two dimensions may lead to ambiguity when multiple occurrences of the same formula are present, as is the case for the formula $\bot\tn\bot$ in the above examples. Where necessary, this will be resolved by adding the names of chosen $\bot$-formulae and $1$-formulae to a diagram. When discussing different proof nets over the same sequent, the sequent will be laid out in a consistent manner.

Figure~\ref{fig:_*_,1,1,1,_*_} uses the diagrammatic notation to display the rewiring relation over the proof nets for the sequent $\bot\tn\bot,1,1,1,\bot\tn\bot$ (an example also highlighted in \cite{Strassburger-Lamarche-2004}). There are 24 proof nets in total (under the restriction that jumps only connect to $1$-occurrences), forming two equivalence classes, each consisting of a single cyclic rewiring path.

\begin{figure}
\[
\begin{array}{@{}ccccccccccc@{}}
	\netA 1223 &\perm*& \netA 1323 &\perm*& \netA 1321 &\perm*& \netA 2321 &\perm*& \netA 2331 &\perm*& \netA 2131 \\ \\ \perm* &&&&&&&&&& \perm* \\ \\
	\netA 1213 &\perm*& \netA 3213 &\perm*& \netA 3212 &\perm*& \netA 3112 &\perm*& \netA 3132 &\perm*& \netA 2132 
	\\ \\ \\ \\ \\
	\netA 1232 &\perm*& \netA 1332 &\perm*& \netA 1312 &\perm*& \netA 2312 &\perm*& \netA 2313 &\perm*& \netA 2113 \\ \\ \perm* &&&&&&&&&& \perm* \\ \\
	\netA 1231 &\perm*& \netA 3231 &\perm*& \netA 3221 &\perm*& \netA 3121 &\perm*& \netA 3123 &\perm*& \netA 2123
	\\ \\
\end{array}
\]
\caption{All proof nets for the sequent $\bot\tn\bot,1,1,1,\bot\tn\bot$}
\label{fig:_*_,1,1,1,_*_}
\end{figure}


\section{Encoding numerical constraints}

In this section we will show how numerical constraints may be encoded in \capsabbrev{mll}, a necessary ingredient for our main construction. We will illustrate the ideas with an example from the literature: an encoding of 3-partition into the problem of \capsabbrev{mll} proof search \cite{Kanovich-1992,Lincoln-Winkler-1994}.

The problem of deciding whether a given \capsabbrev{mll} sequent is provable is \capsabbrev{np}-complete. This was shown first for an extension of \capsabbrev{mll} with weakening \cite{Lincoln-Mitchell-Scedrov-Shankar-1992}, and subsequently for unit-free \capsabbrev{mll} \cite{Kanovich-1992} and for unit-only  \capsabbrev{mll} \cite{Lincoln-Winkler-1994}---note that since \capsabbrev{mll} is conservative over both its unit-only and its unit-free fragment, each of the latter two implies \capsabbrev{np}-completeness for provability in full \capsabbrev{mll}.

Both Kanovich \cite{Kanovich-1992} and Lincoln and Winkler \cite{Lincoln-Winkler-1994} use a reduction from (a variant of) the 3-partition problem to show  \capsabbrev{np}-hardness. Such reductions provide a good illustration of how the linearity of \capsabbrev{mll} can be used to encode combinatoric problems. In this section we will discuss a similar reduction from 3-partition.

To encode simple numerical properties in \capsabbrev{mll}, our primary tools will be formulae of the form $\bot\tn\dotso\tn\bot$ and $1\pr\dotso\pr1$. In particular, two formulae $A$ and $B$ of the former kind may form a proof net with one $C$ of the latter kind if and only if the total number of $\bot$-occurrences in $A$ and $B$ exceeds the number of $1$-occurrences in $C$ by 1: each 1-occurrence must be linked to by a $\bot$, and one additional $\bot$ is needed to connect $A$ and $B$. We will capture this more generally by counting the number of $\bot$-occurrences in $A$ as $i+1$, so that if $j+1$ is the number of $\bot$-occurrences in $B$, the number of $1$-occurrences in $C$ must be $i+j+1$. Below left is an illustration for $i=2$ and $j=3$. In the following, we will introduce an abbreviated notation, illustrated below right. 
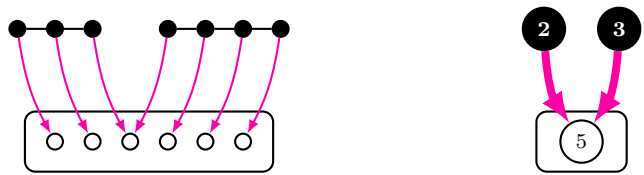
\[
\vc{
  \begin{tikzpicture}[x=5mm,y=-5mm,octo]
	\foreach \x/\i in {0/a,1/b,2/c,4/d,5/e,6/f,7/g} {\node[bullet] (\i) at (\x,0) {};}
	\foreach \i in {1,...,6} {\node[circ] (\i) at (\i,3) {};}
	\draw (a)--(b)--(c) (d)--(e)--(f)--(g);
	\draw[rounded corners] ($(1) + (-.8,-.8)$) rectangle ($(6) + (.8,.8)$);
	\begin{scope}[thepink,->]
			\draw[bend right=10] (a) to (1);
			\draw[bend right=10] (b) to (2);
			\draw[bend right=10] (c) to (3);
			\draw[bend left=10]  (d) to (3);
			\draw[bend left=10]  (e) to (4);
			\draw[bend left=10]  (f) to (5);
			\draw[bend left=10]  (g) to (6);
	\end{scope}
	\node[big_] (A) at (14,0) {$2\ap$};
	\node[big_] (B) at (16,0) {$3\ap$};
	\node[big1] (!) at (15,3) {$5\ap$};
	\draw[rounded corners] ($(!) + (-1.2,-.8)$) rectangle ($(!) + (1.2,.8)$);
	\begin{scope}[thepink,->,big>]
		\draw[bend right=15] (A) to (!);
		\draw[bend left=15]  (B) to (!);
	\end{scope}
  \end{tikzpicture}
}
\]
By $A^n$ we will denote the sequent consisting of $n$ occurrences of a formula $A$. Given a sequent $\Gamma=A_1,\dotsc,A_n$ we will write $\bigtn\Gamma$ for $A_1\!\tn\cdots\tn A_n$, and $\bigparr\Gamma$ for $A_1\!\pr\cdots\pr\! A_n$. In diagrammatic notation, a big disc labelled $n\ap$ will represent the formula $\bigtn(\bot^{n+1})$, and a big circle labelled $n\ap$ will represent the sequent $\1^{n+1}$. A formula $\bigparr(\1^{n+1})$ is represented by a disc enclosed in a box. This extends our diagrammatic notation as follows.

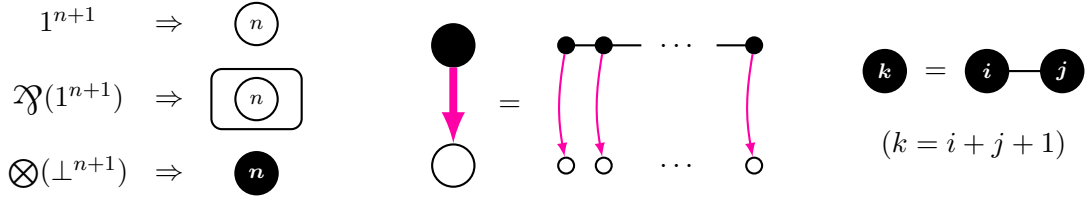
\[
\begin{array}{ccc}
	\1^{n+1}			&\Rightarrow& \bignodes1 \\[10pt]
	\bigparr(\1^{n+1})	&\Rightarrow& \bignodes2 \\[10pt]
	\bigtn(\bot^{n+1})  &\Rightarrow& \bignodes3 \\[10pt]
\end{array}
\qquad\qquad
\vc{
\begin{tikzpicture}[x=5mm,y=8mm,octo]
	\node[big_] (A) at (-1,2) {};
	\node[big1] (!) at (-1,0) {};
	\node at (.5,1) {$=$};
	\draw (2,2) node[bullet] (a) {} -- (3,2) node[bullet] (b) {} -- (4,2) (6,2) -- (7,2) node[bullet] (c) {}; 
	\node[circ] (1) at (2,0) {}; \node[circ] (2) at (3,0) {}; \node[circ] (3) at (7,0) {};
	\node at (5,2) {$\dots$}; \node at (5,0) {$\dots$};
	\begin{scope}[thepink,->]
			\draw[big>] (A) to (!);
			\draw[bend right=10] (a) to (1);
			\draw[bend right=10] (b) to (2);
			\draw[bend right=10] (c) to (3);
	\end{scope}
\end{tikzpicture}}
\qquad\quad
\begin{array}{c}
	\vc{\begin{tikzpicture}[x=5mm,y=5mm,octo] \node[big_] {$k\ap$}; \end{tikzpicture}} ~=~
	\vc{\begin{tikzpicture}[x=5mm,y=5mm,octo] \draw (0,0) node[big_] {$i\ap$} -- (2,0) node[big_] {$j\ap$}; \end{tikzpicture}}
	\\ \\ (k=i+j+1)
\end{array}
\]
A linking between formulae $\bigtn(\bot^n)$ and a sequent $\1^n$ will be represented by a wide arrow, as illustrated above center. For the moment, we will not distinguish between the $n!$ different ways in which such a linking can be made, but when the distinction becomes relevant we will make an explicit choice. Above right, to link a formula $\bigtn(\bot^{k+1})$ to several other formulae, we may display the formula as two discs according to the given arithmetic, representing the formula $\bigtn(\bot^{i+1})\tn~\bigtn(\bot^{j+1})$.

The primary intention behind the chosen notation is to make the arithmetic of connecting formulae $\bigtn(\bot^n)$ to formulae $\bigparr(\1^n)$ apparent in the illustrations. Specifically, the following is a proof net if and only if $n=i_1+\dotso+i_k$.
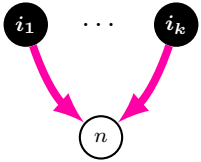
\[
  \vc{\begin{tikzpicture}[x=5mm,y=-5mm,octo]
	\node[big_] (A) at (-2,0) {$i_1\ap$};
	\node[big_] (B) at ( 2,0) {$i_k\ap$};
	\node[big1] (!) at ( 0,3) {$n\ap$};
	\node at (0,0) {$\dots$};
	\begin{scope}[thepink,->,big>]
		\draw[bend right=15] (A) to (!);
		\draw[bend left=15]  (B) to (!);
	\end{scope}
  \end{tikzpicture}}
\]


\subsection{Encoding 3-Partition}

An instance of the 3-partition problem is given by a multiset $I=\{i_1,\dotsc,i_{3n}\}$ of $3n$ natural number values with total sum $\sum I=nk$. The decision problem asks whether a partitioning exists of $I$ into $n$ triples $(i_a,i_b,i_c)$ each with sum $k$. It may be assumed that each $i\in I$ lies between $\nicefrac k4$ and $\nicefrac k2$, so that any partitioning into \emph{subsets} with sum $k$ is a solution, since each subset of sum $k$ necessarily has $3$ elements. The 3-partition problem is strongly \capsabbrev{np}-complete \cite{GareyAndJohnson}, which means it remains \capsabbrev{np}-complete with a unary encoding of the values in $I$.

We will give a simple encoding of 3-partition in MLL proof search, similar to those of Kanovich \cite{Kanovich-1992} and Lincoln and Winkler \cite{Lincoln-Winkler-1994}. A 3-partition instance $I$ is encoded by a sequent consisting of:
\begin{itemize}
	\item a formula $\bigtn(\bot^{i+1})$ for each $i\in I$, and
	\item the formula $\bigtn(K^n)$ where $K=\bigparr(\1^{k+1})$.
\end{itemize}
A solution to the 3-partition problem is encoded as a proof net as follows. If the numbers $a$, $b$, and $c$ are assigned to a triple corresponding to a formula $K_j$, then the corresponding formulae have $a+1$, $b+1$, and $c+1$ occurrences of $\bot$, and $k+1=a+b+c+1$ occurrences of $\1$, respectively. Then as described previously, there is a proof net for these formulae, obtained by attaching each jump to a distinct $\1$, with the exception of three jumps, one each from $a$, $b$, and $c$, overlapping on exactly one instance.

The construction is illustrated below, where the numbers $i$ are conveniently ordered to give the solution $(i_1,i_2,i_3)$, $(i_4,i_5,i_6)$, etc.

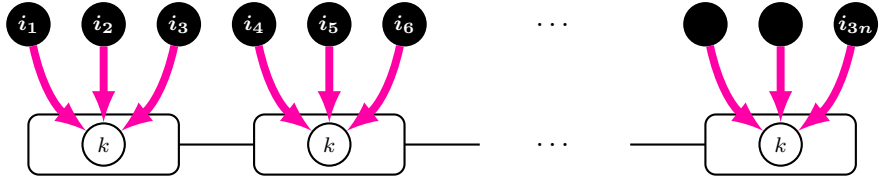
\[
\begin{tikzpicture}[x=10mm,y=8mm,octo]
	\node[big1] (a) at (2,0) {$k\ap$};
	\node[big1] (b) at (5,0) {$k\ap$};
	\node[big1] (c) at (11,0) {$k\ap$};
	\draw[rounded corners] (1,-.5) rectangle (3,.5);
	\draw[rounded corners] (4,-.5) rectangle (6,.5);
	\draw[rounded corners] (10,-.5) rectangle (12,.5);
	\draw (3,0)--(4,0) (6,0)--(7,0) (9,0)--(10,0);
	\node at (8,0) {$\dots$}; \node at (8,2) {$\dots$};
	\foreach \i in {1,...,6} {
		\node[big_] (\i) at (\i,2) {$i\ap_{\i}$};
	}
	\node[big_] (7) at (10,2) {};
	\node[big_] (8) at (11,2) {};
	\node[big_] (9) at (12,2) {$i\ap_{3n}$};
	\begin{scope}[thepink,->,big>]
	\foreach \i/\a in {1/a,4/b,7/c} \draw (\i) to[bend right=20] (\a);
	\foreach \i/\a in {2/a,5/b,8/c} \draw (\i) -- (\a);
	\foreach \i/\a in {3/a,6/b,9/c} \draw (\i) to[bend left=20] (\a);
	\end{scope}
\end{tikzpicture}
\]

The correctness condition ensures that every proof net on a sequent encoding a 3-partition problem gives a solution to that problem. If two jumps from a formula encoding a number $i$ connect to different formulae $K_j$ and $K_m$, a cycle is created immediately. Thus, the jumps from each $i$ must all connect to the same $K$, and the number of occurrences of $\bot$ and $\1$ in the encoding then ensures that when the formulae for $i_a$, $i_b$, and $i_c$, connect to one formulae $K$, their values sum to $k$, $i_a+i_b+i_c=k$.

The encoding of 3-partition serves to illustrate how simple numerical constraints can be encoded in proof nets: by matching up a given number of $\bot$-occurrences and $\1$-occurrences and by separating formulae of the form $\bigparr(\1^n)$ by a tensor. This idea will form the basis of our encoding of non-deterministic constraint logic.


\section{Rewiring proof nets}
\label{sec:rewiring}

In this section we will explore the global rewiring behaviour of proof nets. We will look at notions of subnets; we will introduce a notion of relative \emph{parity} between nets, which if odd, guarantees inequivalence; and we will give a simple account of equivalence for the fragment of \capsabbrev{mll} that omits the par.

The notions and results introduced in this section will be used in the main proofs of the paper, in Section~\ref{sec:correctness}, which show that the encoding of \capsabbrev{ncg}-reconfiguration in \capsabbrev{mll} proof equivalence is correct.


\subsection{Subnets}

We will discuss (and adapt) some convenient standard notions for \capsabbrev{mll} proofs and proof nets, and relate them to rewiring. Firstly we will look at subnets---see also \cite{Bellin-vandeWiele-1995}.

\begin{defi}
A \emph{sub-sequent} $\Delta\leq\Gamma$ of a sequent $\Gamma$ is a sequent consisting of disjoint subformulae of $\Gamma$, preserving names.
\end{defi}

\begin{defi}
A \emph{subnet} $(\Delta,\links') \leq (\Gamma,\links)$ of a proof net is a net such that $\Delta\leq\Gamma$ and $\links'$ is the restriction of $\links$ to the names in $\Delta$.
\end{defi}

The \emph{ports} of a sub-sequent $\Gamma'$ or subnet $(\Gamma',\links')$ are the root vertices of $\Gamma'$. For a vertex $v$ naming a par, tensor, or bottom, the subnets of which it is a port correspond to the possible subproofs of the rule introducing $v$ in a sequentialisation (the subproof of a $\1$-subformula must always be empty).

In the graph of a proof net, a chosen subnet for a par can be made explicit as a \emph{box}, as illustrated below left. Boxes may replace the switching condition as a correctness criterion: in the example, both the outside and the inside of the box form a tree. To make this precise, we will consider the action of \emph{closing} a box, which means it is regarded as a single vertex in the graph, as illustrated below right.
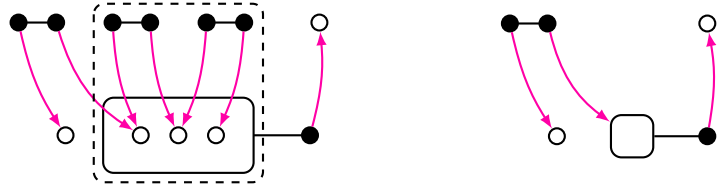
\[
\begin{tikzpicture}[x=5mm,y=-5mm,octo]
	\draw (-2.25,0) node[bullet] (x) {} -- (-1.25,0) node[bullet] (y) {}
		  (.25,0) node[bullet] (a) {} -- (1.25,0) node[bullet] (b) {}
		  ( 2.75,0) node[bullet] (c) {} -- (3.75,0) node[bullet] (d) {};
	\node[circ] (0) at (5.75,0) {};
	\node[circ] (1) at (-1,3) {};
	\node[circ] (2) at (1,3) {}; \node[circ] (3) at (2,3) {}; \node[circ] (4) at (3,3) {};
	\draw[rounded corners] (0,2) rectangle (4,4);
	\draw (4,3) -- (5.5,3) node[bullet] (e) {};
	\begin{scope}[thepink,->]
			\draw[bend right=10] (x) to (1);
			\draw[bend right=20] (y) to (2);
			\draw[bend right=10] (a) to (2);
			\draw[bend right=10] (b) to (3);
			\draw[bend left=10]  (c) to (3);
			\draw[bend left=10]  (d) to (4);
			\draw[bend right=10] (e) to (0);
	\end{scope}
	\draw[dashed,rounded corners] (-.25,-.5) rectangle (4.25,4.25);
\end{tikzpicture}
\qquad\qquad\qquad
\begin{tikzpicture}[x=5mm,y=-5mm,octo]
	\draw (-2.25,0) node[bullet] (x) {} -- (-1.25,0) node[bullet] (y) {};
	\node[circ] (0) at (3,0) {};
	\node[circ] (1) at (-1,3) {};
	\node[draw,rounded corners,minimum size=16pt] (box) at (1,3) {};
	\draw (box)--(3,3) node[bullet] (e) {};
	\begin{scope}[thepink,->]
			\draw[bend right=10] (x) to (1);
			\draw[bend right=20] (y) to (box);
			\draw[bend right=10] (e) to (0);
	\end{scope}
	\path (0,4.25)--(1,4.25);
\end{tikzpicture}
\]

\newcommand\bx{s}

\begin{defi}
A \emph{boxing} $\bx$ for a linking $\links$ for $\Gamma$ assigns a sub-sequent $\bx(v)\leq\Gamma$ to each par-vertex $\named v\pr$ such that 1) $v$ is a port of $\bx(v)$ and 2) boxes are either disjoint or strictly nested: if $\bx(v)\cap\bx(w)\neq\varnothing$ then $\bx(w)<\bx(v)$ or $\bx(v)<\bx(w)$.
\end{defi}

\noindent
In the graph for $\links$ and $\Gamma$, a box $\bx(v)$ may be \emph{closed} by replacing the subgraph over $\bx(v)$ by the single vertex $v$, and replacing every arc into $\bx(v)$ by one onto $v$. For each box $\bx(v)$ we define the \emph{local graph} to be that formed by the subgraph over $\bx(v)$ where each immediately smaller box $\bx(w)<\bx(v)$ is closed. The following is then a variation on the local retraction algorithm by Danos \cite{Danos-1990}.

\begin{prop}
\label{prop:scoping correctness}
A linking $\links$ for $\Gamma$ is a proof net if and only if it has a boxing $\bx$ such that each local graph is a tree.
\end{prop}

\proof
Given a boxing $\bx$, it follows by induction on the nesting of boxes that the graph over each $\bx(v)$ satisfies the switching condition. In the other direction, given a sequentialisation of $(\Gamma,\links)$, a box $\bx(v)\leq\Gamma$ for each $\named v\pr$ is found by taking the conclusion $\Delta, A\named v\pr B$ of its introduction rule, below.
\[
	\infer[\MLLlabel p]{\Delta,A\named v\pr B}{\Delta,A,B}
\]
\qed

In a proof net, the \emph{kingdom} and the \emph{empire} of a vertex $v$ are respectively the smallest and largest subnet that have $v$ as a port. In working with the rewiring relation, the notion of empire can be particularly useful.

\begin{defi}
The \emph{empire} $\emp v$ of a vertex $v$ in $(\Gamma,\links)$ is the largest subnet $(\Delta,\links')$ of which $v$ is a port.
\end{defi}

\begin{prop}[{\cite[Proposition 2.b]{Bellin-vandeWiele-1995}}]
\label{prop:empire propagation}
The empire $\emp v$ is determined by propagation from $v$:
\begin{enumerate}
	\item\label{case:empire propagation 1} through links;
	\item up towards subformulae;
	\item into a tensor if one of its subformulae is in $\emp v-\{v\}$;
	\item into a par if all its subformulae are in $\emp v-\{v\}$.
\end{enumerate}
\end{prop}

\noindent
It should be noted that the above characterisation of  empires relies on the restriction that we have imposed, that jumps target $\1$-formulae only. Otherwise, the first case of the definition should be specialised so that propagation does not traverse jumps into $v$. 

The following three lemmata will show how empires are connected to rewiring. Firstly, a jump from $\named v\bot$ may be rewired to exactly those $\1$-occurences that are in the empire of $v$ (Lemma~\ref{lem:rewiring within empire}). Secondly, rewiring a jump from $\named v\bot$ preserves the empire of $v$, up to that rewiring (Lemma~\ref{lem:rewiring preserves empire}). Thirdly, rewiring the jump from $v$ may add subformulae to the empire of another vertex $w$, or remove subformulae from it, but not both (Lemma~\ref{lem:rewiring affects empires}).

\begin{lem}
\label{lem:rewiring within empire}
For a proof net $(\Gamma,\links)$ where $\links(a)=v$, and $w$ names a $\1$-occurrence in $\Gamma$, the following are equivalent:
\begin{enumerate}
	\item
$\links\perm*\links[\jumpsto aw]$
	\item
$w$ is in the empire $\emp a$; and
	\item
in any switching graph for $(\Gamma,\links)$, the path $\path vw$ does not pass through $a$.
\end{enumerate}
\end{lem}

\proof
By \cite[Proposition 2.a]{Bellin-vandeWiele-1995} 2 and 3 are equivalent.

Next, it is shown that 2 implies 1. The empire $\emp a$ corresponds to the largest subproof $\Sigma$, with as conclusion the introduction rule of $\named a\bot$, in any sequentialisation $\Pi$ of $\links$. By Definition~\ref{def:proofs to nets}, in the translation of $\Sigma$ to a net, $a$ may link to any $1$-occurrence, including $w$.

Finally, it is shown that 1 implies 3, by contraposition. If for some switching of $\links$ the path $\path vw$ passes through $a$, then in $\links[\jumpsto aw]$ there is no path $\path av$ (and two paths $\path aw$) for that switching, so that $(\Gamma,\links[\jumpsto aw])$ is not a net.
\qed

\begin{lem}
\label{lem:rewiring preserves empire}
If $\links\perm*\links[\jumpsto vw]=\links'$ then $\emp v\perm*\emp*v$.
\end{lem}

\proof
Since $v$ may rewire to exactly the same $\1$-occurrences in $\links$ as in $\links'$, by Lemma~\ref{lem:rewiring within empire} the empires $\emp v$ and $\emp*v$ contain the same $\1$-subformulae. That they also share any other subformula $A$ follows by the observation that $A\tn\1$ may replace $A$: by Proposition~\ref{prop:empire propagation} the new $\1$ is in a given empire if and only if $A$ is (unless $v$ names $A$, but in this case $A$ is included in both $\emp v$ and $\emp*v$).
\qed

\begin{lem}
\label{lem:rewiring affects empires}
If $\links\perm*\links'$ where $\emp v$ is a net for the sequent $\Delta$, and $\emp*v$ a net for $\Delta'$, then $\Delta\leq\Delta'$ or $\Delta\geq\Delta'$.
\end{lem}

\proof
Let $\links'=\links[\jumpsto aw]$, where $\links(a)=u$ and $\links'(a)=w$. We will distinguish five cases, depending on three factors: 1) whether $\emp v$ contains $u$; 2) if so, whether $\emp v$ is propagated from $u$ to $a$ or from $a$ to $u$ in case~\ref{case:empire propagation 1} of Proposition~\ref{prop:empire propagation}; and 3) whether $\emp v$ contains $w$.
\begin{itemize}
	\item
If $\emp v$ is propagated from $a$ to $u$, then it includes $\emp a$ as a subnet, because the latter is generated by propagation from $a$. Since $\emp a=\emp*a$ by Lemma~\ref{lem:rewiring preserves empire}, then in $\links'$ the empire of $v$ is propagated from $a$ to $w$, and includes $\emp*a$. It follows that $\Delta=\Delta'$.

	\item
If $\emp v$ is propagated from $u$ to $a$, and also contains $w$, then $\emp v$ has been propagated to $w$ without passing through $a$; otherwise, a switching cycle would be formed by the two propagation paths from $a$ to $w$, one generated by $\emp a$ via $u$, and one generated by $\emp v$, not via $u$. Then $\emp*v$ is propagated from $v$ to $w$ to $a$, and $\Delta=\Delta'$.
 
	\item
If $\emp v$ is propagated from $u$ to $a$, but does not contain $w$, then $\Delta\geq\Delta'$.

	\item
If $\emp v$ contains $w$ but not $u$, then $\Delta\leq\Delta'$.

	\item
If $\emp v$ contains neither $u$ nor $w$, then it also does not contain $a$. Then also $\emp*v$ contains neither $u$, $w$, nor $a$, and $\Delta=\Delta'$.\qed
\end{itemize}



\subsection{Parity}

The linearity of \capsabbrev{mll} means that in a proof or proof net, there is always a certain balance to the number of $\bot$- and $\pr$-occurrences. This observation gives a well-known necessary condition for the provability of a sequent.

\begin{defi}
The \emph{balance} of a sequent is the number of $\bot$s minus the number of $\pr$s and commas.
A sequent is \emph{balanced} if its balance is zero.
\end{defi}

\begin{prop}
\label{prop:unbalanced then uninhabited}
An unbalanced sequent is uninhabited.
\end{prop}


Here, we will introduce a similar necessary condition for the equivalence of two proof nets. Consider the example of the identity and twist maps on $\bot\tn\bot$, below. The two maps are semantically distinct; hence they are often given as a minimal example showing the necessity of having jumps in proof nets.
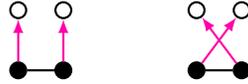
\[
	\twistid0 \qquad\qquad \twistid1
\]
In this case, it is easily seen that neither jump can be rewired, and that the two nets are thus inequivalent; to change from one net to the other, two jumps would need to be rewired simultaneously. This prompts the question of what would happen in a larger net: would a complicated series of rewirings be able to exchange two jumps configured like those of the above nets?

It turns out that this is not the case. Most provable sequents have at least two equivalence classes of proofs, such that for each proof in one class there is a corresponding proof, equal up to the exchange of two jumps, in the other. An example of this is Figure~\ref{fig:_*_,1,1,1,_*_}.

We will capture this idea as follows: we shall associate a \emph{parity} with any pair of linkings $\links$ and $\links'$ over the same sequent $\Gamma$, which may be \emph{even} or \emph{odd}, and we shall find that the parity of equivalent linkings is always even.


For this argument we will work with $n$-ary connectives $\tn$ and $\pr$, and alternating formulae, i.e.\ every argument of a $\tn$ is a $\pr$ and vice versa. The units are given by the $0$-ary connectives, and we need not rule out unary ones. We will consider a given named sequent $\Gamma$, but will assume that it consists of a single formula, if necessary by introducing a $\pr$ at the root.

To be able to compare arbitrary proof nets over $\Gamma$, we will use the following naming scheme for the edges of a switching graph, for any proof net over $\Gamma$. A tensor $\bigtn(A_1,\dotsc,A_n)$ named $v$ has $n$ edges, which we shall name $v(1)$ through $v(n)$; a par $\named v\pr$ has one switched edge, to be named $v(1)$; and $\named v\bot$ has the jump named $v(1)$. The naming scheme identifies edges across all switching graphs of all proof nets for $\Gamma$.

Taking a different perspective, we may alternatively consider a switching graph as a directed tree, rooted in the root connective of $\Gamma$. This establishes a bijection between the non-root vertices and the edges, which associates each vertex with the edge connecting it to its parent.


\begin{figure}
\[
	\vc{\parityNetA} \qquad\qquad\qquad \vc{\parityNetB}
\]
\caption{A switching graph of a proof net as a directed tree}
\label{fig:directed switching graph}
\end{figure}

The example in Figure~\ref{fig:directed switching graph} displays a proof net on the left, and on the right the switching graph choosing the edge $\edge ij$ for the par $i$, and the edge $\edge rg$ for the root par $r$. The induced bijection associates for example the vertex $g$ and the edge named $r(1)$, the the vertex $h$ and the edge $g(1)$, and the vertex $a$ and the edge $h(1)$; it further associates $f$ with $f(1)$, and $c$ with $c(3)$.

Given two proof nets $\links$ and $\links'$ for $\Gamma$, and a switching graph for each (not necessarily given by the same switching of $\Gamma$), we obtain two bijections between edges and (non-root) vertices. Composing these gives a permutation on the non-root vertices.

\begin{defi}
The \emph{parity} of two switching graphs for proof nets $\links$ and $\links'$ for a sequent $\Gamma$ is the parity of their induced permutation.
\end{defi}

\noindent
We will show that both 1) rewiring and 2) choosing a different switching induce even parity. By 2) we may define the parity of two proof nets $\links$ and $\links'$ to be that over arbitrary switching graphs; then by 1) it follows that proof nets with odd parity are inequivalent.

We will demonstrate 1), while 2) is similar. Let $\links\perm*\links'$ by rewiring a jump $\jump va$ to $\jump vb$. By fixing a switching for $\Gamma$, we obtain a switching graph for each of the two nets, where the jump is named $v(1)$ in each. There are two possibilities, illustrated below. On the left, if the jump $v(1)$ is directed upward, then the target of each edge in the directed switching graphs remains the same---in particular $v(1)$ has target $v$---and the induced permutation is the identity.

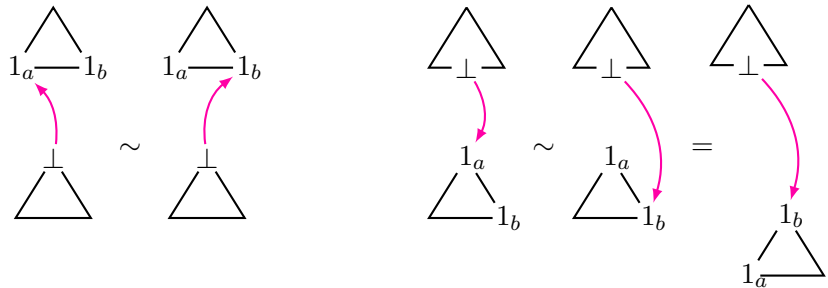
\[
	\vc{\parityNetC1} ~\perm~ \vc{\parityNetC2}
\qquad\qquad\qquad
	\vc{\parityNetD1} ~\perm~ \vc{\parityNetD2} ~\kern5pt{=}\kern-5pt~ \vc{\parityNetD3}
\]

Above on the right, if the jump from $v$ is directed downward, the subtree of $a$ will get the new root node $b$. Then the vertices that are associated with a new edge are exactly those on the path $\path ab$ in the switching graph, as illustrated below.
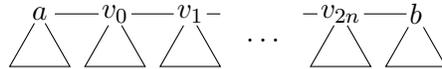
\[
	\parityNetE
\]
Since the connectives in $\Gamma$ were assumed to be strictly alternating, there are an odd number of vertices on this path, $2n+3$: each even $v_i$ must be a $\bot$ or $\pr$, while each odd $v_j$ must be a $\1$ or $\tn$. The permutation induced is then as follows. Since $v$ has target $a$ in the first switching graph, and target $b$ in the second, it takes $a$ to $b$. Further, since an edge connecting $v_i$ and $v_{i+1}$ has target $v_{i+1}$ in the first, but target $v_i$ in the second graph, the permutation takes $v_{i+1}$ to $v_i$. The complete permutation is then a cyclic one taking each vertex on the path $\path ab$ to the previous, and the first to the last. A cyclic permutation of odd length has even parity.

The above argument gives us 1), that rewiring has even parity. To see that the same argument also gives 2), it is sufficient to consider that choosing a different switching for a single par is essentially the same operation as rewiring, if the par is considered a $\bot$ and the switched edge a jump. We may thus conclude that:

\begin{prop}
Two equivalent proof nets have even parity.
\end{prop}


For our encoding, in Section~\ref{sec:encoding}, this poses the following challenge (the \emph{parity problem}): when we want the encoding to produce proof nets that are equivalent, we must ensure that these do not inadvertently have odd parity.




\begin{figure*}
\[
\begin{array}{ccccc}
	\doubleExchange uvxy &\quad\perm*\quad&
	\doubleExchange uyxy &\quad\perm*\quad&
	\doubleExchange uyxu
	\\ \\[3pt] &&&& \perm*
	   \\[3pt] &&&& \doubleExchange vyxu
	\\ \\[3pt] &&&& \perm*
	   \\[3pt] 
	\doubleExchange vuyx &\perm*&
	\doubleExchange vuyu &\perm*&
	\doubleExchange vyyu
\end{array}
\]
\caption{Double exchange of jumps (Lemma~\ref{lem:double exchange})}
\label{fig:double exchange}
\end{figure*}

\subsection{Equivalence without \texorpdfstring{$\protect\pr$}{par}}

Let a \emph{basic} sequent be one of formulae constructed only over $\1$, $\bot$, and $\tn$. After removing dangling $\bot$-formulae and replacing subformulae $\1\tn A$ with $A$, basic sequents consist of formulae of the form $1$ or $\bigtn(\bot^n)$ with $n\geq2$. Provability for basic sequents is entirely determined by balance:

\begin{prop}
A balanced basic sequent is inhabited.
\end{prop}

We will show that, similarly, equivalence for basic sequents is determined by parity. An immediate observation is that a proof net for a basic sequent with only one tensor-formula (a formula of the form $\bigtn(\bot^n)$ for $n\geq 2$), every $\1$ is linked to by exactly one jump, which means that no rewiring is possible.

\begin{prop}
A basic sequent $\1^n,\bigtn(\bot^n)$ is inhabited by $n!$ inequivalent proof nets.
\end{prop}

In the following we will characterise equivalence for basic sequents with two or more tensor-formulae.

\begin{lem}
\label{lem:double exchange}
Let $\links$ be a proof net with for every switching a path
\[
	\links(a)\mathrel{\xpmuj}a\mathrel{\xpath}b\mathrel{\xjump}\links(b)\mathrel{\xpath}\links(c)\mathrel{\xpmuj}c\mathrel{\xpath}d\mathrel{\xjump}\links(d)~.
\]
Then $\links\perm\links[\jumpsto a{\links(b)}][\jumpsto b{\links(a)}][\jumpsto c{\links(d)}][\jumpsto d{\links(c)}]$.
\end{lem}

\proof
By the rewiring path shown in Figure~\ref{fig:double exchange}.
\qed

\begin{lem}
\label{lem:level0 max binary}
A basic sequent with at least two tensor-formulae has at most two equivalence classes of proof nets.
\end{lem}

\proof
By induction on the size of a sequent $\Gamma$. The base case is the sequent $\1,\1,\1,\bot\tn\bot,\bot\tn\bot$, which has two equivalence classes of proof nets, shown in Figure~\ref{fig:_*_,1,1,1,_*_}. For the inductive step, let $\Gamma=\Gamma',A\tn\named a\bot,\named z\1$ where $A\tn\named a\bot$ is a largest $\tn$-formula in $\Gamma$. It will be shown that any net $\links$ is equivalent to one $\links'$ where $\links'(a)=z$; then by induction, the subnet $\links'$ restricted to $\Gamma',A$ belongs to one of two equivalence classes. To find $\links'$, there are two cases.

1) The path $\path az$ is via $\links(a)$. If $\links(a)=z$, we are done. Otherwise, by Lemma~\ref{lem:rewiring within empire} $\links'$ may be obtained from $\links$ by changing only $\links'(a)=z$.

2) The path $\path az$ is via some $\named b\bot$ in $A$. Firstly, if $\links(b)\neq z$, use Lemma~\ref{lem:rewiring within empire} to re-attach $b$ to $z$. Next, let $\named c\bot$ and $\named d\bot$ be occurrences in a separate formula $B$ such that $c$ links to the same $\1$-occurrence as some $\bot$ in $A$. Then $\links'$ is obtained by linking $c$ to $b$, and applying Lemma~\ref{lem:double exchange} to exchange the targets of $a$ and $b$, as well as those of $c$ and $d$.
\qed

\begin{prop}
\label{prop:parity determines equivalence}
For a basic sequent with at least two tensor-formulae, two proof nets with even parity are equivalent.
\end{prop}

\proof
By Lemma~\ref{lem:level0 max binary} the sequent has at most 2 equivalence classes. Given two proof nets of even parity, both must be in the other equivalence class than a proof net with odd relative parity to both, which exists by exchanging two jumps from one tensor-formula.
\qed


\section{Constraint logic}
\label{sec:ncl}

\emph{Non-deterministic constraint logic} (NCL) \cite{Hearn-Demaine-2005,Demaine-Hearn-2008,GamesPuzzlesAndComputation} is a simple graph-rewriting formalism, introduced as a convenient tool for \capsabbrev{pspace}-hardness reduction. It is exactly for this purpose that we shall use it.

The graphs of NCL, called \emph{constraint graphs}, have directed, weighted edges, and a rewrite step (or \emph{move}) consists of reversing the direction of exactly one edge. Each vertex in a constraint graph has an \emph{inflow constraint}, a natural number value, and moves are required to preserve the property that the \emph{inflow} at each vertex, the sum weight of its incoming edges, never drops below the inflow constraint (i.e.\ the inflow constraint dictates a \emph{minimum} inflow).

The specific problem we will use is the \emph{configuration--to--configuration} problem, which asks whether a path of rewrite steps exists between two constraint graphs. This problem is \capsabbrev{pspace}-complete, and remains so under various restrictions: inflow constraints can be fixed at the value 2, edge weights can be restricted to the value 1 and 2, the graphs can be required to be planar, and loops and double edges (edges connecting the same pair of vertices) can be prohibited, for example.

We will need no such restrictions, and as they would not significantly simplify our encoding, we will work with the general case. However, for simplicity of exposition, in the examples of constraint graphs in this section we will use the following conventions. Edges will have weight 1 or 2; those of weight 1 are drawn as thin red arrows, and those of weight 2 as thicker blue arrows. Vertices with inflow constraint 2 are drawn as grey circles. For convenience, we will assume that dangling edges are connected to vertices of weight zero, which are omitted from illustrations. These zero-inflow vertices can be seen as ``connection points'', where the graph may be connected to a wider network of graphs.

An example rewrite sequence is given in Figure~\ref{fig:NCL example}. This particular constraint graph encodes a logical conjunction, and is not unlike an \textsc{and}-gate: in order to invert the weight 2 edge on the right, both weight 1 edges on the left must be inverted first.

\begin{figure}
\[
	\andexample 1 \quad\perm*\quad
	\andexample 2 \quad\perm*\quad
	\andexample 3 \quad\perm*\quad
	\andexample 4
\]
\caption{A series of reconfiguration steps in a constraint graph}
\label{fig:NCL example}
\end{figure}

In the formal definition, a constraint graph is an \emph{undirected} graph, for which a \emph{configuration} assigns an orientation to each edge. It is useful for us to generalise the notion of configuration a little: we will allow \emph{partial} configurations, where the direction of edges may be left undefined, as long as the inflow constraints are satisfied by the directed edges. We will use $\star$ to indicate an undefined value, to avoid overloading $\bot$.

\begin{defi} 
A \emph{constraint graph} $G=(V,E,c,v,w)$ consists of:
\begin{itemize}
	\item $V$ a set of vertices,
	\item $E$ a set of edges,
	\item $c\colon V\to\mathbb N$ an \emph{inflow constraint} on each vertex,
	\item $v\colon E\to\mathcal P(V)$ where $|v(e)|\in\{1,2\}$ for $e\in E$, a set of 1 or 2 vertices for each edge,
 	\item $w\colon E\to\mathbb N$ a \emph{weight} on each edge.
\end{itemize}
\end{defi}

\begin{defi}
A \emph{(partial) configuration} for a constraint graph $G=(V,E,c,v,w)$ is a (partial) function $\gamma\colon E\to V$ such that
\begin{itemize}
	\item
for every edge $e$, $\gamma(e)\in v(e)\cup\{\star\}$,
	\item
for every vertex $v$,
\[
	c(v) \leq \sum\{w(e)\mid\gamma(e)=v\}~.
\]
\end{itemize} 
\end{defi}

In the above definition, the first condition states that the weight of an edge may be assigned only to one of its vertices, or left undefined, but not to any other vertex in the constraint graph. The second condition states that the total weight of the incoming edges of a vertex must be at least its inflow constraint.

We will write $\gamma[e\mapsto v]$ for the partial configuration that directs $e$ to $v$, and any other edge $e'$ to $\gamma(e')$. An edge that may be reversed will be called \emph{mobile}; formally, an edge $e$ is \emph{mobile} in $\gamma$ if $\gamma[e\mapsto\star]$ is a partial configuration.

\begin{defi}
A \emph{reconfiguration step} $(\perm*)$ relates two partial configurations for $G$ that differ in value (or definedness) on exactly one edge:
\[
	\gamma~\perm*~\gamma[e\mapsto u]\qquad\text{ if }\gamma(e)\neq u\text{ and }u\in v(e)\cup\{\star\}~.
\]
The reflexive--transitive closure of $(\perm*)$ will be denoted $(\perm)$.
\end{defi}

\emph{Non-deterministic constraint graph reconfiguration} or \emph{\capsabbrev{ncg}-reconfiguration} is the problem of deciding whether two total configurations of a constraint graph are connected by a sequence of reconfiguration steps. More formally, an instance of \capsabbrev{ncg}-reconfiguration is a triple $(G,\gamma,\delta)$ consisting of a constraint graph $G$ and two configurations $\gamma$ and $\delta$ for $G$. The decision problem then asks whether $\gamma\perm\delta$.

This is the problem shown to be \capsabbrev{pspace}-complete by Hearne and Demaine \cite{GamesPuzzlesAndComputation}:

\begin{thm}[\cite{GamesPuzzlesAndComputation}, Theorem~5.15]
\capsabbrev{ncg}-reconfiguration is \capsabbrev{pspace}-complete.
\end{thm}

To show that this result extends to our setting, with partial configurations, we have the following proposition.

\begin{prop}
\label{prop:partial simulates total reconfiguration}
For total configurations $\gamma$ and $\delta$, if $\gamma\perm\delta$ then $\gamma$ and $\delta$ are also connected by a sequence of reconfiguration steps over total configurations only.
\end{prop}

\proof
By the following two observations: firstly, 
if $\gamma\perm*\delta$ for partial configurations, then these may be completed to total configurations $\gamma'\perm*\delta'$ or $\gamma'=\delta'$; and secondly,
if $\gamma'$ and $\gamma''$ are total configurations that both agree with a partial configuration $\gamma$ where the latter is defined, then $\gamma'$ and $\gamma''$ are connected by reversing the edges on which they disagree one after another.
\qed

\begin{figure}
\[
\begin{array}{ccccccccc}
	\boat0 &\perm& \boat1 &\perm& \boat2 &\perm& \boat3 &\perm& \boat4
	\\ \\ &&&&&&&& \perm \\ \\
	\boat9 &\perm& \boat8 &\perm& \boat7 &\perm& \boat6 &\perm& \boat5
\end{array}
\]
\caption{A reconfiguration path requiring two inversions of a single edge}
\label{fig:boats}
\end{figure}

\begin{figure}
\[
	\boats
\]
\caption{Nesting graphs to grow reconfiguration paths exponentially}
\label{fig:big boat}
\end{figure}

The \capsabbrev{pspace}-completeness of \capsabbrev{ncg}-reconfiguration may seem surprising at first. To give a feel for the dynamics that make this problem \capsabbrev{pspace}-hard, we shall exhibit an example where an exponentially-long reconfiguration sequence is needed. The existence of such rewiring sequences is an essential property for the hardness result: if reconfiguration sequences could be given a polynomial bound, they would be polynomial-time witnesses, placing the problem within NP.

The example is then as follows. Figure~\ref{fig:boats} gives a reconfiguration sequence, clockwise from the top left, on a constraint graph with two connection points, top and bottom. As such, the graph operates as would a single edge from top to bottom, and the reconfiguration sequence inverts the direction of this would-be edge. The key point is that in doing so, the central horizontal edge must be inverted \emph{twice}. Then by nesting the graph within itself, where one graph replaces the central edge of another, as illustrated in Figure~\ref{fig:big boat}, a graph is created with an edge that requires 4 inversions. Each added level of nesting doubles the number of inversions that the central edge must make, while growing the graph by only a constant factor. Thus, the length of a rewiring path may grow exponentially with the size of the graph.


\section{Encoding constraint logic}
\label{sec:encoding}

We will demonstrate the \capsabbrev{pspace}-hardness of \capsabbrev{mll} proof equivalence by an encoding of \capsabbrev{ncg}-reconfiguration in \capsabbrev{mll} proof nets. A sequent will encode a constraint graph, and a proof net a configuration, such that rewiring  the proof net corresponds to reconfiguration.

The natural basis of the encoding is the use of jumps to indicate where the weight of an edge is allocated. The main problem is then how to restrict the movement of such jumps, so that they may assign weight only to the vertices of one particular edge, and not to any other vertices. The central design idea of our encoding is to use the combinatorics of connecting formulae of the form $\bigtn(\bot^n)$ to formulae $\bigparr(1^m)$ to solve this problem.

The two basic components of our encoding will be \emph{weight elements}, representing the weight value of edges, and \emph{constraint elements}, which encode the inflow constraint of vertices. For now, we will work under the assumption that each unit of weight on an edge is encoded by exactly one weight element---later, we will amend this to use multiple weight elements for each unit of weight. A weight element will be a construction over a number of $\bot$-formulae, whose jumps may connect to the $\1$-formulae of a constraint element.

Below top is a weight element, below bottom a constraint element.

\[
\begin{array}{ccc}
	  \vc{\begin{tikzpicture}[x=5mm,y=5mm,octo]
	 	\node[big_] (A) at (0,0) {$i\ap$}; 
	 	\node[big_] (B) at (2,0) {$j\ap$};
		\node[big_] (C) at (4,0) {$k\ap$};
		\draw[rounded corners] (-1,-1) rectangle (5,1);
	  \end{tikzpicture}}
	&=&
	  \bigtn(\bot^{i+1})~~\pr~~\bigtn(\bot^{j+1})~~\pr~~\bigtn(\bot^{k+1})
\\ \\
	  \vc{\begin{tikzpicture}[x=5mm,y=5mm,octo]
		\node[big1] (1) at (0,0) {$m\ap$}; 
		\node[big1] (2) at (3,0) {$n\ap$};
		\draw[rounded corners] (-1,-1) rectangle (1,1) (2,-1) rectangle (4,1);
		\draw (1,0)--(2,0);
	  \end{tikzpicture}}
	&=&
	  \bigparr(\1^{m+1})~\tn~\bigparr(\1^{n+1})
\end{array}
\]

\noindent
For all edges and vertices in the encoding of a constraint graph, the sum $i+j+k=m+n$ will be the same---this way, a priori any weight element may connect to any constraint element. The value $m$ will be unique for each vertex, and used for all constraint elements in the encoding of that vertex. The weight element above is for an edge connecting the two vertices for which $m=i$ and $m=i+j$: with those constraint elements it is possible to form a proof net, as illustrated below.

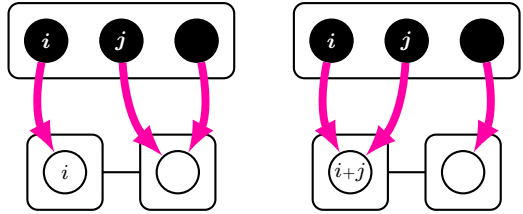
\[
	\connectingElements 1 \qquad \connectingElements 2
\]

\noindent
To ensure that a weight element may not form a proof net with a constraint element of another vertex, the values of $m$ and $n$ are chosen such that $m\equiv1$ and $n\equiv2 \pmod 3$, and accordingly $i\equiv1$, $j\equiv0$, and $k\equiv2\pmod 3$. It then becomes impossible for $m$ to equal any other sum over $i$, $j$, and $k$, so that forming a proof net requires either $m=i$ or $m=i+j$.

Several further constructions are used in our encoding. Firstly, in a constraint graph, the sum of all weights is usually greater than the sum of all inflow constraints---otherwise, no edge can be mobile, or no configuration exists. Correspondingly, an encoding will have weight elements not connected to constraint elements. These will instead connect to additional, separate $\1$-formulae, referred to as \emph{weight absorbers}, as shown below. A weight element that is connected to absorbers will be called \emph{free}.
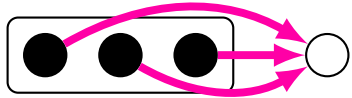
\[
\begin{tikzpicture}[x=5mm,y=5mm,octo]
	\node[big_] (A) at (0,0) {}; 
	\node[big_] (B) at (2,0) {};
	\node[big_] (C) at (4,0) {};
	\draw[rounded corners] (-1,-1) rectangle (5,1);
	\node[big1] (1) at (7.5,0) {};
	\begin{scope}[thepink,->,big>]
		\draw[bend left=30] (A) to (1);
		\draw[bend right=30] (B) to (1);
		\draw (C) to (1);
	\end{scope}
\end{tikzpicture}
\]

An edge in a constraint graph will be encoded by an \emph{edge-gadget}, illustrated below left, constructed by stringing together a number of similar weight elements, plus a single $\bot$-formula which we will refer to as the \emph{indicator}. Illustrated below right is a \emph{vertex-gadget} encoding a vertex, formed by a number of constraint elements plus a single \emph{indicator target}.

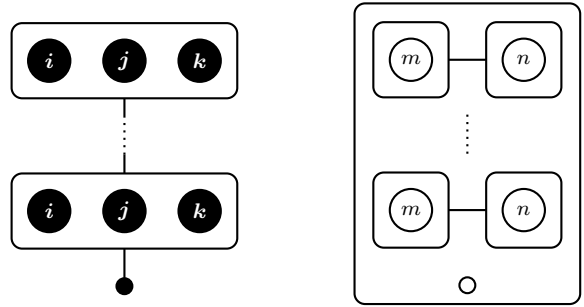
\[
\begin{tikzpicture}[x=5mm,y=5mm,octo]
	\node[big_] (A) at (0,0) {$i\ap$}; 
	\node[big_] (B) at (2,0) {$j\ap$};
	\node[big_] (C) at (4,0) {$k\ap$};
	\draw[rounded corners] (-1,-1) rectangle (5,1);
	\node[big_] (D) at (0,4) {$i\ap$}; 
	\node[big_] (E) at (2,4) {$j\ap$};
	\node[big_] (F) at (4,4) {$k\ap$};
	\draw[rounded corners] (-1,3) rectangle (5,5);
	\draw (2,1)--(2,1.5) (2,2.5)--(2,3);
	\draw[dotted] (2,1.5)--(2,2.5);
	\draw (2,-1)--(2,-2) node[bullet] {};
	\path (0,-2.5)--(1,-2.5);
\end{tikzpicture}
\qquad\qquad
\begin{tikzpicture}[x=5mm,y=5mm,octo]
	\node[big1] (1) at (0,0) {$m\ap$}; 
	\node[big1] (2) at (3,0) {$n\ap$};
	\draw[rounded corners] (-1,-1) rectangle (1,1) (2,-1) rectangle (4,1);
	\draw (1,0)--(2,0);
	\node[big1] (3) at (0,4) {$m\ap$};
	\node[big1] (4) at (3,4) {$n\ap$};
	\draw[rounded corners] (-1,3) rectangle (1,5) (2,3) rectangle (4,5);
	\draw (1,4)--(2,4);
	\draw[dotted] (1.5,1.5)--(1.5,2.6);
	\node[circ] at (1.5,-2) {};
	\draw[rounded corners] (-1.5,-2.5) rectangle (4.5,5.5);
\end{tikzpicture}
\]

The natural way for edge-gadgets to link up with a vertex-gadget is shown in Figure~\ref{fig:connect example}: indicators connect to indicator targets, and weight elements connect to constraint elements or weight absorbers. The illustration depicts two equivalent proof nets, both encoding the same (fragment of a) constraint graph: a single vertex that is the target of a weight-1 edge (on the left, in red) and a weight-2 edge (on the right, in blue). The inflow of the vertex is 3, and its constraint is 2; then 2 weight elements are connected to the constraint elements in the vertex-gadget, while one remains free (connected to weight absorbers). It is essential to the functioning of the encoding that any re-distribution of weight elements gives an equivalent proof net, as illustrated by the example. Then in the second proof net, the single weight element of the left edge-gadget is free, which means the indicator jump can be rewired to connect to a different vertex.

In the encoding, the indicator jumps determine which vertex is the target of the edge, and while they may in principle assign any vertex, the weight elements of the edge-gadget can only form proof nets with the constraint elements of the correct vertices. An edge-gadget will be called \emph{free} if all its weight elements are free. The vertex-gadgets in an encoding will all be gathered in a $\tn$-formula, which means they are connected, so that free edge-gadgets may attach their indicator to any of them.

\begin{figure}

\noindent$\CONexample 2{.60}\hfill\perm\hfill\CONexample 1{.60}$

\caption{Two edge-gadgets connecting to one vertex-gadget}
\label{fig:connect example}
\end{figure}

In Figure~\ref{fig:big example} we give a complete encoding of the example rewiring sequence shown in Figure~\ref{fig:NCL example}, a reconfiguration sequence for a constraint graph functioning as an \textsc{and}-gate. The given proof nets are all equivalent, and their corresponding constraint graphs are shown near the bottom of the figure. The central vertex-gadget encodes the central constraint-2 vertex in the graph, while its three implicit constraint-0 vertices, at the dangling ends of the three edges, are encoded by the three formulae connected to the central gadget.

\newcommand\ANDmark[1]{\makebox[0pt][r]{\Large{\textbf{#1}}}}

\begin{figure}[p]
\[
\begin{array}{@{}r@{\hspace{16pt}}|@{\hspace{16pt}}l@{}}
	\ANDexample 1 \ANDmark 1 & \ANDexample 2 \ANDmark 2 \\ \\ \hline \\
	\ANDexample 3 \ANDmark 3 & \ANDexample 4 \ANDmark 4 \\ \\ \hline \\
	\ANDexample 5 \ANDmark 5 & \ANDexample 6 \ANDmark 6 \\ \\ \hline
\end{array}
\]
\vspace{20pt}
\[
\begin{array}{ccccccc}
	\andexample 1 &\perm& \andexample 2 &\perm& \andexample 3 &\perm& \andexample 4 \\
	\textbf{1} && \textbf{2} && \textbf{3~4~5} && \textbf{6}
\end{array}
\]
\caption{Encoding an \textsc{and}-gate in proof nets}
\label{fig:big example}
\end{figure}

\subsection{Weight adjustment}

So far we have assumed that one weight element may encode one unit of weight in the constraint graph. However, there is a minor issue that prevents this straightforward approach. Although one weight element cannot `fill' an inappropriate constraint element, two weight elements \emph{can}, in the way illustrated below. In such a situation, the weight element that should have been linked to this constraint element becomes free, and the edge it belongs to may inappropriately become free, too. 
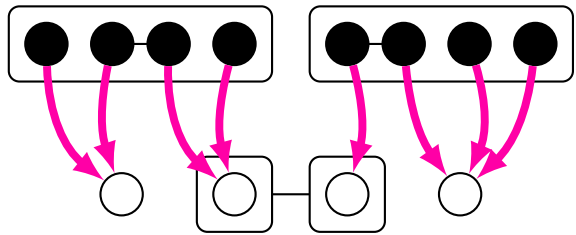
\[
\begin{tikzpicture}[x=5mm,y=-5mm,octo]
	\node[big_] (A) at (-1,0) {}; 
	\draw (.75,0) node[big_] (B1) {} -- (2.25,0) node[big_] (B2) {};
	\node[big_] (C) at (4,0) {};
	\draw[rounded corners] (-2,-1) rectangle (5,1);
	\draw (7,0) node[big_] (D1) {} -- (8.5,0) node[big_] (D2) {}; 
	\node[big_] (E) at (10.25,0) {};
	\node[big_] (F) at (12,0) {};
	\draw[rounded corners] (6,-1) rectangle (13,1);
	\node[big1] (1) at (1,4) {};
	\node[big1] (2) at (4,4) {}; 
	\node[big1] (3) at (7,4) {};
	\node[big1] (4) at (10,4) {};
	\draw[rounded corners] (3,3) rectangle (5,5) (6,3) rectangle (8,5);
	\draw (5,4)--(6,4);
	\begin{scope}[thepink,->,big>]
		\draw[bend right=25] (A)  to (1);
		\draw[bend right=15] (B1) to (1);
		\draw[bend right=25] (B2) to (2);
		\draw[bend right=15] (C)  to (2);
		\draw[bend left=15] (D1) to (3);
		\draw[bend right=15] (D2) to (4);
		\draw[bend left=20] (E)  to (4);
		\draw[bend left=20] (F)  to (4);
	\end{scope}
\end{tikzpicture}
\]
To resolve this issue, it is sufficient to increase the number of weight elements used to encode one unit of weight. In the following, we will investigate how many are needed.


In an inappropriate linking as shown above, since both halves of the constraint element are connected, the weight elements must be disconnected---otherwise, the switching condition would be violated. That means the weight elements must belong to different edges. Note that it is possible to use more than one weight element from the same edge to fill one half of a constraint element, or even weight elements from different edges, but we are interested here only in the \emph{minimum} number of edges needed to inappropriately fill a vertex-gadget.


As the linkings above illustrate, it may occur that one subformula of a weight element $A\pr B\pr C$ fills one half of a constraint element. In the pathological case for $(v_i)$ where $i$ is very small or very large, a weight element can fill three halves of different constraint elements, as illustrated in Figure~\ref{fig:pathological}. As in the illustration, the other three halves may be filled by weight elements of a different edge---so to fill 3 constraint elements requires 2 inappropriate edges. To fill the next three constraint elements, at most 1 previous inappropriate edge may be used, and one additional one is needed. To fill $3n$ constraint elements inappropriately therefore requires $n+1$ edges. It thus suffices to encode one unit of weight on an edge by $3\times|E|$ weight elements (where $|E|$ is the number of edges in the constraint graph), and correspondingly to encode one unit of inflow constraint by $3\times|E|$ constraint elements.

\begin{figure}
\[
\begin{tikzpicture}[x=-5mm,y=-5mm,octo]
	\node[big1] (1) at (0,3) {};
	\foreach \i in {0,3,6} {
		\node[big_] (A\i) at (3.5,\i) {}; 
		\node[big_] (B\i) at (5,  \i) {}; 
		\node[big_] (C\i) at (7,  \i) {};
		\node[big_] (D\i) at (9,  \i) {};
		\draw (A\i)--(B\i);
		\draw[rounded corners] ($(2.5,-1) + (0,\i)$) rectangle ($(10,1) + (0,\i)$);
		\node[big1] (2\i) at (13,\i) {}; 
		\node[big1] (3\i) at (16,\i) {};
		\draw[rounded corners] ($(12,-1) + (0,\i)$) rectangle ($(14,1) + (0,\i)$);
		\draw[rounded corners] ($(15,-1) + (0,\i)$) rectangle ($(17,1) + (0,\i)$);
		\draw (14,\i)--(15,\i);
		\begin{scope}[thepink,->,big>]
			\draw[bend left=30] (B\i) to (2\i);
			\draw[bend left=15]  (C\i) to (2\i);
			\draw (D\i) to (2\i);
		\end{scope}
	}
	\begin{scope}[thepink,->,big>]
		\draw[bend right=20] (A6) to (1);
		\draw                (A3) -- (1);
		\draw[bend left=20]  (A0) to (1);
	\end{scope}
	\draw (6.25,1) -- +(0,1) (6.25,4) -- +(0,1);
	\draw[rounded corners] (11.5,-1.5) rectangle (17.5,7.5);
	\foreach \i in {1,3,5} {
		\node[big_] (E\i) at (20,  \i) {};
		\node[big_] (F\i) at (21.5,\i) {};
		\draw (E\i)--(F\i);
	}
	\draw[rounded corners] (19,0) rectangle (22.5,6);
	\node[big1] (4) at (24,3) {};
	\begin{scope}[thepink,->,big>]
		\draw[bend right=20] (E5) to (36);
		\draw                (E3) -- (33);
		\draw[bend left=20]  (E1) to (30);
		\draw[bend left=20]  (F5) to (4);
		\draw                (F3) -- (4);
		\draw[bend right=20] (F1) to (4);
	\end{scope}
\end{tikzpicture}
\]
\caption{A weight element connecting to three constraint elements}
\label{fig:pathological}
\end{figure}


\subsection{The complete encoding}

The complete encoding of a constraint graph $G$ will then be a sequent $\Gamma$ consisting of:
\begin{enumerate}
 	\item
all vertex-gadgets, combined in a single formula via tensors, 
	\item
all edge-gadgets as individual formulae, and
	\item
a sufficient number of weight absorbers ($\1$-formulae).
\end{enumerate}
A configuration for $G$ will be encoded as a proof net for $\Gamma$, and conversely each proof net for $\Gamma$ may be interpreted as a (partial) configuration for $G$. The encoding will be made formal in the next section.


\section{Formalising the encoding}
\label{sec:correctness}

We will formalise the encoding of \capsabbrev{ncg}-reconfiguration into \capsabbrev{mll} proof equivalence that was informally introduced in the previous section. A constraint graph $G$ will be encoded as a sequent $\itn G$, and a configuration $\gamma$ for $G$ will be encoded as a proof net $\itn\gamma$ for $\itn G$. We will show that $\gamma\perm\delta$ if and only if $\itn\gamma\perm\itn\delta$ (modulo a small adjustment to ensure even parity between $\itn\gamma$ and $\itn\delta$).

For a constraint graph $G=(V,E,c,v,w)$, let $|V|$ and $|E|$ denote the number of vertices and edges, and let $|c|$ and $|w|$ denote the sum of all inflow constraints and the sum of all weights, respectively:
\[
	|c| = \sum_{v\in V}c(v) \qquad\qquad |w| = \sum_{e\in E}w(e)~.
\]

\begin{defi}
\label{def:graph encoding}
The \emph{encoding} $\itn G$ of a constraint graph $G$ is a sequent constructed as follows. 
Let $G=(V,E,c,v,w)$ with $|V|=n$, $|E|=m$, $V=\{v_1,\dotsc,v_n\}$, and $E=\{e_1,\dotsc,e_m\}$.

\noindent
The encoding of a vertex $v_k$ is the formula
\[
	\itn{v_k} = \bigparr\big(C(k,n)^{3m\,\times\,c(v_k)} \big)\pr\1
\]
where each \emph{constraint element} $C(k,n)$ is the formula
\[
	C_n(k) = \bigparr\big(1^{3k+2}\big) \tn \bigparr\big(1^{3(n-k)+3}\big)~.
\]

\noindent
The encoding of an edge $e$ connecting vertices $v_i$ and $v_j$ with $i\leq j$ is the formula
\[
	\itn{e} = \bigtn\big(W(i,j,n)^{3m\,\times\,w(e)}\big)\tn\bot
\]
where each \emph{weight element} $W(i,j,n)$ is the formula
\[
	W(i,j,n) = \bigtn\big(\bot\!^{3i+2}\big)\pr\bigtn\big(\bot\!^{3(j-i)+1}\big)\pr\bigtn\big(\bot\!^{3(n-j)+3}\big)\,.
\]

\noindent
The encoding of the graph $G$ is the sequent
\[
	\itn G = \itn{v_1}\tn\dotso\tn\itn{v_n}, \itn{e_1},\dotsc,\itn{e_m}, 1^p
\]
where $p=3m\times(|w|-|c|)\times(3n+4)$.

\end{defi}

In the above definition, the final $\1$-subformula of a vertex-gadget $\itn{v_k}$ is its \emph{indicator target}; the final $\bot$-subformula of an edge-gadget $\itn{e}$ is its \emph{indicator}; and in the completed encoding $\itn G$ the $p$ instances of $\1$ are the \emph{weight absorbers}.
In a constraint graph $G$, a vertex $v$ and an edge $e$ will be called \emph{appropriate} (for each other) if $v\in v(e)$, and \emph{inappropriate} otherwise.
This notion is extended to vertex-gadgets $\itn v$ and edge-gadgets $\itn e$ in $\itn G$, and their respective constraint elements and weight elements.

A configuration $\gamma$ for a constraint graph $G$ will be encoded as a proof net for the sequent $\itn G$.
Firstly we will define a standard way of linking a weight element to a constraint element.

\begin{defi}
\label{def:standard linkings}
For $W=W(i,j,n)=X\pr Y\pr Z$ a weight element,
\begin{enumerate}
	\item
for a constraint element $C=C(i,n)=A\tn B$, the \emph{standard linking} for the sequent $C,W$ links the first $\bot$ in $X$ to the first $\1$ in $A$, the first $\bot$ in $Y$ and $Z$ each to the first $\1$ in $B$, and each remaining $\bot$ in $X,Y,Z$ to a remaining $\1$ in $A,B$ in their order of occurrence;
	\item
for $C=C(j,n)=A\tn B$, the \emph{standard linking} for $C,W$ is defined as above, except the first $\bot$ in $Y$ links to the first $\1$ in $A$;
	\item
the \emph{standard linking} for the sequent $W,\1^{3n+4}$ links the first $\bot$ in $X$, $Y$, and $Z$ to the first $\1$, and each remaining $\bot$ to a remaining $\1$ in order of occurrence.
\end{enumerate}
\end{defi}

\noindent The standard linkings defined in the second and third case of the above definition are illustrated below.

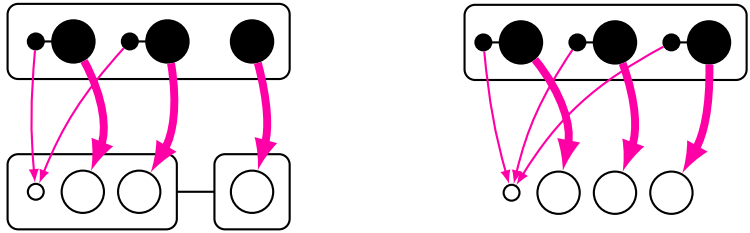
\[
\begin{tikzpicture}[x=5mm,y=5mm,octo]
	\node[circ] (1) at (-.25,0) {};
	\node[big1] (2) at (1,0) {};
	\node[big1] (3) at (2.5,0) {};
	\node[big1] (4) at (5.5,0) {};
	\draw[rounded corners] (-1,-1) rectangle (3.5,1) (4.5,-1) rectangle (6.5,1);
	\draw (3.5,0)--(4.5,0);
	\draw (-.25,4) node[bullet] (a) {} -- (.75,4) node[big_] (A) {};
	\draw (2.25,4) node[bullet] (b) {} -- (3.25,4) node[big_] (B) {};
	\node[big_] (C) at (5.5,4) {};
	\draw[rounded corners] (-1,3) rectangle (6.5,5);
	\begin{scope}[thepink,->]
		\draw[bend right=5] (a) to (1);
		\draw[bend right=10] (b) to (1);
		\draw[big>,bend left=25] (A) to (2);
		\draw[big>,bend left=20] (B) to (3);
		\draw[big>,bend left=15] (C) to (4);
	\end{scope}
\end{tikzpicture}
\qquad\qquad\qquad
\begin{tikzpicture}[x=5mm,y=5mm,octo]
	\node[circ] (1) at (.75,0) {};
	\node[big1] (2) at (2,0) {};
	\node[big1] (3) at (3.5,0) {};
	\node[big1] (4) at (5,0) {};
	\draw ( 0  ,4) node[bullet] (a) {} -- ( 1  ,4) node[big_] (A) {};
	\draw ( 2.5,4) node[bullet] (b) {} -- ( 3.5,4) node[big_] (B) {};
	\draw ( 5  ,4) node[bullet] (c) {} -- ( 6  ,4) node[big_] (C) {};
	\draw[rounded corners] (-.5,3) rectangle (7,5);
	\path (0,-1) -- (1,-1);
	\begin{scope}[thepink,->]
		\draw[bend right=5] (a) to (1);
		\draw[bend right=10] (b) to (1);
		\draw[bend right=15] (c) to (1);
		\draw[big>,bend left=25] (A) to (2);
		\draw[big>,bend left=20] (B) to (3);
		\draw[big>,bend left=15] (C) to (4);
	\end{scope}
\end{tikzpicture}
\]

\begin{prop}
\label{prop:element linkings}
Standard linkings are proof nets.
\end{prop}

The encoding of a configuration is then as follows.

\begin{defi}
\label{def:configuration encoding}
The \emph{encoding} $\itn\gamma$ of a total configuration $\gamma$ for a constraint graph $G$ is a linking $\links$ for $\itn G$, constructed incrementally for each successive edge $e$ and for each successive weight element $W$ within $e$, as follows.
Let $\gamma(e)=v$; firstly, the indicator of $\itn e$ is linked to the indicator target of $\itn v$.
Then successively for each weight element $W$ in $e$, if $\itn v$ has a first free constraint element $C$, extend $\itn\gamma$ to include the standard linking on $C,W$; otherwise, extend $\itn\gamma$ by the standard linking on the sequent consisting of $W$ plus the first $3n+4$ free weight absorbers.
\end{defi}

\begin{prop}
If $\gamma$ is a total configuration for $G$ then $\itn\gamma$ is a proof net for $\itn G$.
\end{prop}

\proof
Using Proposition~\ref{prop:scoping correctness}, it is sufficient to give a suitable box for each $\pr$. 
The box of each weight element $W$ is the sequent $C,W$ or $W,1^{3n+4}$ of its standard linking, which forms a proof net by Proposition~\ref{prop:element linkings}.
The box of each vertex-gadget $\itn v$ contains the edge-gadgets $\itn e$ such that $\gamma(e)=v$, plus all the weight absorbers within boxes of weight elements inside $\itn e$.
Since the weights of the connected edges $e$ sum to more than the inflow constraint of $v$, there are no unused constraint elements remaining in $\itn v$.
After closing the box of each $W$, each edge-gadget in the box of $\itn v$ becomes a single string of connected vertices, connected to other edge-gadgets only via the indicator target of $\itn v$, thus forming a tree.
\qed

Finally, we will encode an instance of the NCL reconfiguration problem, consisting of an NCL graph $G$ with two configurations $\gamma$ and $\delta$, for which the problem asks whether $\gamma\perm\delta$. Here we need to ensure that the two encodings as proof nets have even parity; otherwise, they will never be equivalent. We have opted for a simple and straightforward solution: first we encode both configurations as $\itn\gamma$ and $\itn\delta$; then we check their parity, and if it is odd, we adjust it by swapping two weight absorbers on $\itn\delta$.

\begin{defi}
The \emph{encoding} of an instance of NCL-reconfiguration $(G,\gamma,\delta)$ is the triple $(\itn G,\itn\gamma,\itn\delta_\gamma)$ where
\[
	\itn\delta_\gamma=\left\{
	 \begin{array}{ll}\itn\delta & \text{if $\itn\gamma$ and $\itn\delta$ have even parity} \\
	 				  \itn\delta' & \text{otherwise}\end{array}\right.
\]
and $\itn\delta'$ is $\itn\delta$ with the first two weight absorbers $a$ and $b$ exchanged: $v\mathbin{\xjump}a$ in $\itn\delta'$ if and only if $v\mathbin{\xjump}b$ in $\itn\delta$, and vice versa, while $\itn\delta'$ agrees with $\itn\delta$ on any other jump.
\end{defi}

In the remainder, we will show that our encoding is correct, i.e.\ that $\gamma\perm\delta$ if and only if $\itn\gamma\perm\itn\delta_\gamma$. This will be separated into two parts: completeness ($\Rightarrow$) and soundness ($\Leftarrow$).


\subsection{Completeness}

Given a reconfiguration path $\gamma\perm\delta$ over total configurations, we will demonstrate a rewiring sequence between $\itn\gamma$ and $\itn\delta_\gamma$.
The central part of the argument will be to show how the weight element linking to a constraint element may be exchanged for another (Lemma~\ref{lem:octopus roll}).
Before and after the exchange, the constraint element and the weight element connecting to it will be in a standard linking.
The linking between weight elements and weight absorbers need not be standard: it will be shown that weight absorbers may be freely rearranged, as long as parity remains even (Lemma~\ref{lem:permute weight absorbers}).

For a reconfiguration step $\gamma\perm*\delta$ where the edge $e$ changes direction from $v$ to $w$, the rewiring sequence $\itn\gamma\perm\itn\delta_\gamma$ will be as follows.
First, the weight elements of edge-gadgets connecting to $\itn v$ are rearranged to match their target configuration, in $\itn\delta_\gamma$, which means the weight elements of $\itn e$ connect only to weight absorbers.
Then $\itn e$ is moved from $\itn v$ to $\itn w$ by rewiring its indicator link, from the indicator of $\itn v$ to that of $\itn w$.
Next, the weight elements connecting to $\itn w$ are rewired to match $\itn\delta_\gamma$, and finally, weight absorbers are rearranged to match $\itn\delta_\gamma$ as well.

To describe the intermediate stages of such a rewiring sequence, call an edge-gadget $\itn e$ \emph{well linked} if: 1) its indicator connects to the indicator target of an appropriate vertex-gadget $\itn v$, and 2) each weight element is either in a standard linking with a constraint element of $\itn v$, or is free (linked only to weight absorbers, in arbitrary fashion). In the following main lemma we will exchange the weight element linking to a constraint element. The lemma considers a sequent consisting of just a vertex-gadget, some appropriate edge-gadgets, and sufficiently many weight absorbers.

\begin{lem}
\label{lem:octopus roll}
Let $\links$ be a proof net for $\Gamma=\itn v,\itn{e_1},\dotsc,\itn{e_m},1^p$ such that each edge-gadget $\itn e_i$ is well linked. Let $W_i$ and $W_j$ be weight elements in edge-gadgets $\itn{e_i}$ and $\itn{e_j}$ respectively; let $W_i$ be linked to $C$ in $\itn v$, and let $W_j$ be linked to weight absorbers $1^n$. Then there is a net $\links'\perm\links$ in which $W_j$ is well linked to $C$, $W_i$ is linked to $1^n$, and $\links'$ agrees with $\links$ otherwise.
\end{lem}

\renewcommand\scalefactor{0.80}

\newcommand\displayOcto[1]{%
   \medskip%
   \centerline{%
	 \scale{#1}%
  }%
  \medskip%
}

\proof
Let $W_i=X\pr Y\pr Z$, $W_j=P\pr Q\pr R$, and $C=A\tn B$.
The rewiring path will be illustrated for the case where $X,Y,A$ and $P,A$, and thus also $Z,B$ and $Q,R,B$, are balanced sequents; other cases are similar.

\medskip

\noindent
\textbf{1.}\quad The initial configuration is illustrated below. Other edges, other weight and constraint elements, and the outer box of the vertex-gadget, are omitted. The vertices $i$, $j$, and  $v$ are the indicators of $\itn{e_i}$ and $\itn{e_j}$ and the indicator target of $\itn v$, respectively.

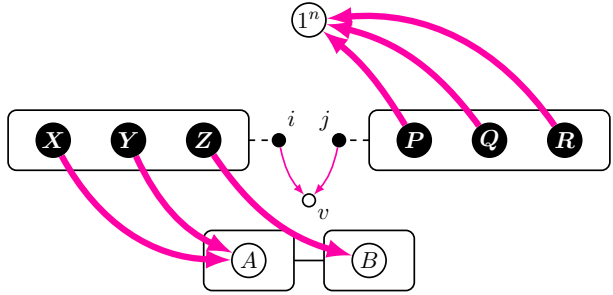
\displayOcto{\octorollA1}

\noindent
\textbf{2.}\quad The jump $\jump iv$ is rewired to a weight absorber, together with only the jumps from the first $\bot$ of each of $P$, $Q$, and $R$:

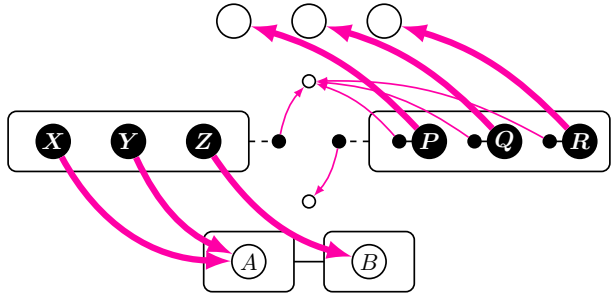
\displayOcto{\octorollB2}

\noindent
\textbf{3.}\quad The jump from the first $\bot$ of $P$ is moved to $A$, and those from the first $\bot$ of $Q$ and $R$ are moved to $B$:

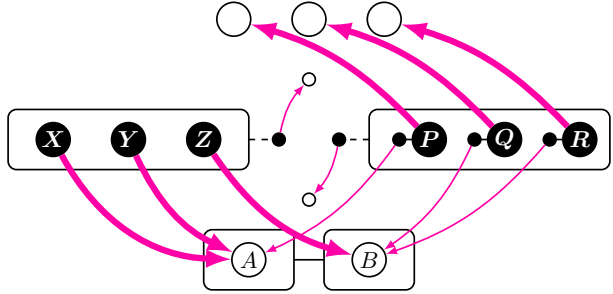
\displayOcto{\octorollB3}

\noindent
\textbf{4.}\quad In the present configuration, the jumps from the weight elements form two subnets: one over the sub-sequent $X,Y,A,P,1^m$, and one over the sub-sequent $Z,B,Q,R,1^k$, for some $m$ and $k$. By Proposition~\ref{prop:parity determines equivalence}, these subnets are equivalent to any other over the same sequent with which they have even parity. It is then sufficient to choose linkings so that $P\pr Q\pr R$ is in a standard linking with $A\tn B$, with a minor adjustment: two jumps from $P$ to $A$ should remain exchanged, compared to the standard linking, for step 6 below. The jumps of $X,Y,Z$ link to the weight absorbers $1^m$ and $1^k$, with one remaining jump from $Y$ to $A$ and one from $Z$ to $B$.

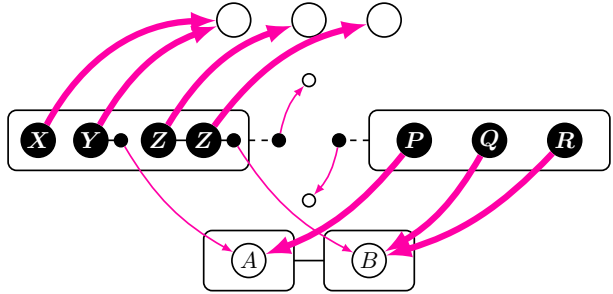
\displayOcto{\octorollC}

\noindent
\textbf{5.}\quad The jump from $Z$ to $B$ is moved towards a weight absorber connected to $X,Y$.

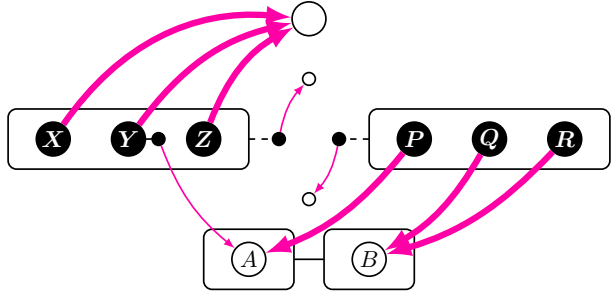
\displayOcto{\octorollD1}

\noindent
\textbf{6.}\quad The jump from $Y$ to $A$ is the one remaining connection between the edge-gadgets $\itn{e_i}$ and $\itn{e_j}$. Lemma~\ref{lem:double exchange} allows to swap the two targets of the jump from $Y$ to $A$ and the jump from $i$, and simultaneously undo the exchange in the two jumps from $P$ to $A$ added in step 4.

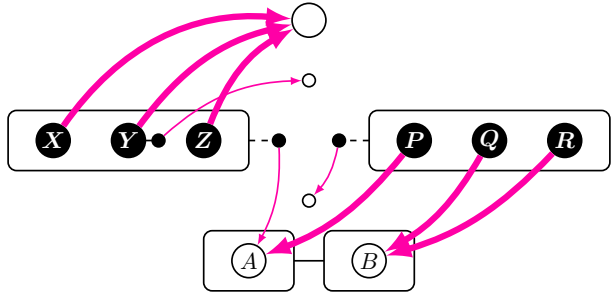
\displayOcto{\octorollD2}

\noindent
\textbf{7.}\quad The jump from $i$ is re-attached to $v$ to yield the final configuration:

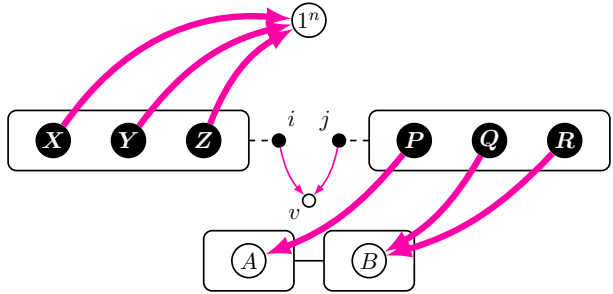
\displayOcto{\octorollA2}

\qed

\noindent In the exchange of weight elements in the above lemma, weight elements link to weight absorbers in arbitrary fashion. To be able to correct for this, the next lemma will demonstrate that, modulo parity, weight absorbers can be re-arranged at will.

\begin{lem}
\label{lem:permute weight absorbers}
If $\links$ and $\links'$ are well-linked proof nets for $\itn G$ that are equal up to an even permutation of weight absorbers, then $\links\perm\links'$ if $\links$ and $\links'$ have at least one free edge-gadget $\itn e$.
\end{lem}

\proof
To reconfigure $\links'$ into $\links$, we will use the double exchange operation of Lemma~\ref{lem:double exchange} to exchange two arbitrary weight absorbers at a time, as well as two chosen jumps in the free edge-gadget $\itn e$.
Firstly, let $e_0$ be the indicator of $\itn e$.
Note that since $\itn e$ is free, $e_0$ may re-attach anywhere within the proof net.

We will need two basic operations: 1) to exchange two arbitrary weight absorbers $v$ and $w$ linked from outside $\itn e$; and 2) to exchange an arbitrary weight absorber linked from inside $\itn e$ with one linked from outside $\itn e$.
Since we are using double exchanges, each operation will also exchange two arbitrary jumps from $\itn e$ (but not that from $e_0$).
By using operation 1) twice, we obtain a third, where four arbitrary weight absorbers linked from inside $\itn e$ are pairwise exchanged.
Since $\links'$ and $\links$ differ by an even permutation of weight absorbers, these three operations together give $\links'\sim\links$.

To perform operation 1), exchanging $v$ and $w$ as well as the jumps from $e_1$ and $e_2$ in $\itn e$, first connect $e_0$ to $v$.
Then apply Lemma~\ref{lem:double exchange} three times: exchanging $v$ and $w$, and the targets of $e_0$ and $e_1$; exchanging $w$ and $v$, and the targets of $e_1$ and $e_2$; and exchanging $v$ and $w$, and the targets of $e_2$ and $e_0$.
The result is a net exchange of $v$ and $w$, and the targets of $e_1$ and $e_2$.

Operation 2) is performed similarly.
In both cases, if one of the weight absorbers exchanged is linked to by multiple $\bot$-occurrences within the same weight element, these may be temporarily attached elsewhere.
\qed

\begin{lem}
\label{lem:completeness}
If $\gamma\perm\delta$ for total configurations $\gamma$ and $\delta$, then $\itn\gamma\perm\itn\delta_\gamma$.
\end{lem}

\proof
By Proposition~\ref{prop:partial simulates total reconfiguration} we may assume that $\gamma\perm\delta$ by a sequence of reconfiguration steps over total configurations.
We will prove that if $\gamma\perm*\delta$ then $\itn\gamma\perm\itn\delta$ or $\itn\gamma\perm\itn\delta'$.
The same proof will show the corresponding case for $\itn\gamma'$ instead of $\itn\gamma$, so that the general case for $\gamma\perm\delta$ follows by transitivity.

Let $\gamma$ and $\delta$ agree on every edge except $e$, where $\gamma(e)=v$ and $\delta(e)=w$.
Firstly, using Lemma~\ref{lem:octopus roll}, for the edges $d$ other than $e$ such that $\gamma(d)=v$, the weight elements of the edge-gadgets $\itn d$ may be linked to the constraint elements of $\itn v$, in accordance with the target configuration $\itn\delta$.
Since $e$ is mobile in $\gamma$, the weights of the edges $d$ suffice to fill the inflow constraint of $v$, and correspondingly the weight elements of edge-gadgets $\itn d$ suffice to fill the constraint elements of $\itn v$, so that $\itn e$ is free.
Next, the indicator vertex of $\itn e$, which links to the indicator target of $\itn v$, is re-attached to the indicator target of $\itn w$.
Again using Lemma~\ref{lem:octopus roll}, the weight elements of edge-gadgets connected to $\itn w$, including $\itn e$, may be linked in accordance with $\itn\delta$.
The resulting proof net is $\itn\delta$ modulo a permutation of weight absorbers; then it is equivalent to either $\itn\delta$ or $\itn\delta'$ by Lemma~\ref{lem:permute weight absorbers}.
\qed


\subsection{Soundness}

It will be shown that proof net rewiring $\itn\gamma\perm\itn\delta_\gamma$ is sound for \capsabbrev{ncg}-reconfiguration, i.e\ that it implies $\gamma\perm\delta$.
To each proof net $\links$ for an encoded constraint graph $\itn G$ we will associate a partial configuration $\gamma=\coitn\links$, such that 1) a rewiring step between $\links$ and $\links'$ corresponds to a reconfiguration path $\coitn\links\perm\coitn{\links'}$, and 2) the function $\coitn-$ is a retraction of the encoding of configurations, $\coitn{\itn\gamma}=\gamma$.

The configuration $\coitn\links$ will assign an edge $e$ to a vertex $v$ when, in the proof net $(\itn G,\links)$, the edge-gadget $\itn e$ is in the empire of the vertex-gadget $\itn v$.
Firstly, it will be shown that $\coitn\links$ is a partial function.

\begin{prop}[{\cite[Proposition 1.i]{Bellin-vandeWiele-1995}}]
\label{prop:tensor has disjoint subnets}
In a proof net, if vertices $v$ and $w$ are joined by a tensor, then any two subnets of which they are respective ports are disjoint.
\end{prop}

\begin{lem}
In a proof net for $\itn G$, an edge-gadget $\itn e$ belongs to the empire of at most one vertex-gadget $\itn v$.
\end{lem}

\proof
Since vertex-gadgets are joined by a tensor, the lemma is immediate from Proposition~\ref{prop:tensor has disjoint subnets}.
\qed

Next, it will be shown that the appropriate edge-gadgets in the empire of a vertex-gadget $\itn v$ contain sufficient weight elements to fill the constraint elements of $\itn v$.

\begin{lem}
\label{lem:appropriate edge weights}
In a proof net for $\itn G$, for each node $v$ in $G$, the weights of the appropriate edge-gadgets in the empire of $\itn v$ are equal to or greater than the constraint of $v$.
\end{lem}

\proof
Let $|E|=m$.
We will show that the inappropriate edge-gadgets in $G$ are insufficient to fill $m$ weight elements in $\itn v$.
Since weight elements come in multiples of $m$, it follows that to fill the $m\times c(v)$ constraint elements of $v$, there must be at least as many appropriate weight elements available.

Let $e$ be inappropriate for $v$; let $C=A\tn B$ be a constraint element in $\itn v$, and $W=X\pr Y\pr Z$ a weight element  in $\itn e$. To find a proof net for the sequent $C,W$ requires to assign balanced boxes to the par-formulae $A$ and $B$. Since $e$ is inappropriate, the sequents $A,X$ and $A,X,Y$ are not balanced, while the balance of each of the other sequents of $A$ with one or more of $X,Y,Z$ is always $1$ or $2\pmod 3$. It follows that there is no proof net for $C,W$.

Next, it will be shown that to balance $m$ constraint elements requires at least $m+1$ inappropriate edge-gadgets.
Two edge-gadgets may balance at most three constraint elements: one weight element $W$ has three subformulae, which may each balance at most one half of a constraint element; the other halves may be balanced by different weight elements of the second edge-gadget.
In the same way, adding one further edge-gadget allows at most three further constraint elements to be filled, since previous edges may connect to only one half of each constraint element.
Then to balance $3m$ constraint elements inappropriately requires $m+1$ edge-gadgets.
\qed

Using the above, a proof net for $\itn G$ may be interpreted as a configuration for $G$.

\begin{defi}
For a proof net $\links$ for $\itn G$, let $\coitn\links$ be the partial configuration for $G$ where
\[
	\coitn\links(e)=
	\left\{\begin{array}{ll}
		v & \text{if $e$ is appropriate for $v$ and $\itn e$ is in the empire of $\itn v$} \\
		\star & \text{otherwise.}
	\end{array}\right.
\]
\end{defi}

To observe that an encoding $\itn\gamma$ decodes back to $\gamma$, note that whenever $\gamma(e)=v$ the indicator of $\itn e$ in $\itn\gamma$ connects to the indicator target of $\itn v$; this is sufficient to place $\itn e$ in the empire of $\itn v$. We have the following proposition.

\begin{prop}
\label{prop:retraction}
$\coitn{\itn\gamma}=\gamma$ and $\coitn{\itn\gamma'}=\gamma$.
\end{prop}

The following lemma then shows that rewiring corresponds to reconfiguration under $\coitn-$.

\begin{lem}
\label{lem:soundness}
If $\links\perm\links'$ are proof nets for $\itn G$ then $\coitn\links\perm\coitn{\links'}$.
\end{lem}

\proof
The proof will consider the case where $\links\perm*\links'$; the general case follows by transitivity.
By Lemma~\ref{lem:rewiring affects empires}, the empire of each vertex-gadget $\itn v$ contains either a subset of, a superset of, or exactly the same edge-gadgets in $\links'$ as it does in $\links$.
Let $\itn {e_1}$ through $\itn {e_n}$ be the edge-gadgets moving into or out of the empire of $\itn v$.
By Lemma~\ref{lem:appropriate edge weights} other edge-gadgets must fill the constraint elements of $\itn v$.
Then the corresponding edges $e_1$ through $e_n$ are mobile in $\coitn\links$.
It follows that $\coitn\links\perm\coitn{\links'}$ by moving each $e_i$ in turn, and repeating the process for other vertices.
\qed


\section{\texorpdfstring{\capsabbrev{mll}}{MLL} proof equivalence is \texorpdfstring{\capsabbrev{pspace}}{PSpace}le.-complete}

We are now ready to state our main theorem.

\begin{thm}
\capsabbrev{mll} proof equivalence is \capsabbrev{pspace}-complete.
\end{thm}

\proof
\capsabbrev{mll} proof equivalence has at most non-deterministic polynomial space complexity: a proof net may be represented in linear space (with respect to a proof); a single rewiring step is performed without requiring additional space; and a non-deterministic algorithm may guess the correct rewiring sequence.
Then by Savitch's Theorem \cite{Savitch-1970} \capsabbrev{mll} proof equivalence is in \capsabbrev{pspace}.

\capsabbrev{pspace}-hardness is by the encoding of \capsabbrev{ncg}-reconfiguration. Recall that an instance of \capsabbrev{ncg}-reconfiguration $(G,\gamma,\delta)$ is encoded by the triple $(\itn G,\itn\gamma,\itn\delta_\gamma)$. Then:
\[
	\gamma\perm\delta \qquad\iff\qquad \itn\gamma\perm\itn\delta_\gamma~.
\]
The direction $(\Rightarrow)$ is by Lemma~\ref{lem:completeness}; the direction $(\Leftarrow)$ is by Lemma~\ref{lem:soundness} and Proposition \ref{prop:retraction}. Encoding an instance of \capsabbrev{ncg}-reconfiguration $(G,\gamma,\delta)$ involves first computing $\itn G$, $\itn\gamma$, and $\itn\delta$, and next computing the parity of $\itn\gamma$ and $\itn\delta$ and adjusting $\itn\delta$ if needed. This can be done in polynomial time in the size of $(G,\gamma,\delta)$.
\qed

\section*{Acknowledgements}
We would like to thank Beniamino Accattoli, Dominic Hughes, and Lutz Stra{\ss}burger for helpful discussions and comments. We would also like to thank the anonymous referees for their insightful remarks.

\bibliographystyle{alpha}
\bibliography{LMCS}
\end{document}